\documentclass[12pt,twoside]{mitthesis}
\usepackage{lgrind}
\usepackage{mathtools}
\usepackage{empheq}
\usepackage{amsfonts}
\usepackage{graphicx}
\usepackage{amsmath}
\usepackage{amssymb}
\usepackage{geometry}
\usepackage{fancyhdr}
\usepackage{url}
\usepackage{epsfig}
\usepackage{epstopdf}
\usepackage{color}
\usepackage{enumerate} \usepackage{framed} \usepackage{multicol} \usepackage{sectsty}
\usepackage{subcaption}
\usepackage{floatrow}

\makeatletter
\def\@seccntformat#1{\@ifundefined{#1@cntformat}%
   {\csname the#1\endcsname}%  default
   {\csname #1@cntformat\endcsname}}%  enable individual control
\newcommand{\section@cntformat}{\thesection.\ \ }
\newcommand{\subsection@cntformat}{\thesubsection\ \ }
\def\thesubsubsection{\Roman{subsubsection}}
\newcommand{\subsubsection@cntformat}{\thesubsubsection. \ \ }
\makeatother

\newtheorem{Def}{Definition}  
 \newtheorem{Rem}{Remark} \newtheorem{Proof}{Proof}
\newtheorem{Prop}{Proposition} 
\newtheorem{Modif}{Modification} \newcommand{\mb}{\mathbf} \newcommand{\mbb}{\mathbb}
 \newcommand{\fa}{\forall} 
\newcommand{\tb}{\textbf} \newcommand{\mc}{\mathcal} \newcommand{\tx}{\text}

\definecolor{dark_purple}{rgb}{0.4, 0.0, 0.4}
\definecolor{mit_red}{rgb}{.6,.1, .2}

\def\QEQ{{%
    \setbox0\hbox{$\heartsuit$}%
    \rlap{\hbox to \wd0{\hss \large$\nearrow$\hss}}\box0
}}

\begin{document}

 % -*-latex-*-
% $Log: cover.tex,v $
% Revision 1.7  2010/04/29 11:35:46  bryt
% changed department chair from Art Smith to Terry Orlando
% changed default copyright flag from author to MIT, left directions for changing it back
% 
% Revision 1.6  1999/10/21 14:49:31  boojum
% changed comment referring to documentstyle
%
% Revision 1.5  1999/10/21 14:39:04  boojum
% *** empty log message ***
%
% Revision 1.4  1997/04/18  17:54:10  othomas
% added page numbers on abstract and cover, and made 1 abstract
% page the default rather than 2.  (anne hunter tells me this
% is the new institute standard.)
%
% Revision 1.4  1997/04/18  17:54:10  othomas
% added page numbers on abstract and cover, and made 1 abstract
% page the default rather than 2.  (anne hunter tells me this
% is the new institute standard.)
%
% Revision 1.3  93/05/17  17:06:29  starflt
% Added acknowledgements section (suggested by tompalka)
% 
% Revision 1.2  92/04/22  13:13:13  epeisach
% Fixes for 1991 course 6 requirements
% Phrase "and to grant others the right to do so" has been added to 
% permission clause
% Second copy of abstract is not counted as separate pages so numbering works
% out
% 
% Revision 1.1  92/04/22  13:08:20  epeisach
\title{
  {Efficiency-Risk Tradeoffs in Dynamic Oligopoly Markets}
  \\ {-- with application to electricity markets}}

\author{Qingqing Huang}
\prevdegrees{B.Eng in Electronic Engineering and BBA in General Business Management}
\department{Department of Electrical Engineering and Computer Science}
% If the thesis is for two degrees simultaneously, list them both
% separated by \and like this:
% \degree{Doctor of Philosophy \and Master of Science}
% \degree{B.Eng Electronic Engineering \and BBA in General Business Management}
\degree{Master of Science in Electrical Engineering and Computer Science}
\degreemonth{June} \degreeyear{2013} \thesisdate{May 22, 2013}

%% By default, the thesis will be copyrighted to MIT.  If you need to copyright
%% the thesis to yourself, just specify the `vi' documentclass option.  If for
%% some reason you want to exactly specify the copyright notice text, you can
%% use the \copyrightnoticetext command.  
%\copyrightnoticetext{\copyright IBM, 1990.  Do not open till Xmas.}

% If there is more than one supervisor, use the \supervisor command
% once for each.
\supervisor{Munther A. Dahleh}{ Professor}

% This is the department committee chairman, not the thesis committee
% chairman.  You should replace this with your Department's Committee
% Chairman.
\chairman{Leslie Kolodziejski}{Chairman, Department Committee on Graduate Students}

% Make the titlepage based on the above information.  If you need
% something special and can't use the standard form, you can specify
% the exact text of the titlepage yourself.  Put it in a titlepage
% environment and leave blank lines where you want vertical space.
% The spaces will be adjusted to fill the entire page.  The dotted
% lines for the signatures are made with the \signature command.
\maketitle

% The abstractpage environment sets up everything on the page except
% the text itself.  The title and other header material are put at the
% top of the page, and the supervisors are listed at the bottom.  A
% new page is begun both before and after.  Of course, an abstract may
% be more than one page itself.  If you need more control over the
% format of the page, you can use the abstract environment, which puts
% the word "Abstract" at the beginning and single spaces its text.

%% You can either \input (*not* \include) your abstract file, or you can put
%% the text of the abstract directly between the \begin{abstractpage} and
%% \end{abstractpage} commands.

% First copy: start a new page, and save the page number.
\newpage
% Uncomment the next line if you do NOT want a page number on your
% abstract and acknowledgments pages.
% \pagestyle{empty}
\setcounter{savepage}{\thepage}
\begin{abstractpage}

%\begin{abstract}

In an abstract framework, we examine how a tradeoff between efficiency and risk arises in
different dynamic oligopolistic markets. We consider a scenario where there is a reliable
resource provider and agents which enter and exit the market following a random process.
Self-interested and fully rational agents can both produce and consume the resource. They
dynamically update their load scheduling decisions over a finite time horizon, under the
constraint that the net resource consumption requirements are met before each
individual's deadline.
%\tc{red}{mention applications of the stylized model?}
  %%

We first examine the system performance under the non-cooperative and cooperative market
architectures, both under marginal production cost pricing of the resource.
%%
% In the non-cooperative market, agents' demands for the resource are strategic
% substitutes, and each agent schedules his consumption to optimize his expected cost of
% implementing his schedule; while in the cooperative market, the agents cooperate in the
% decision making process to optimize aggregate expected cost.
The statistics of the stationary aggregate demand processes induced by the two market
architectures show that although the non-cooperative load scheduling scheme leads to an
efficiency loss - widely known as the ``price of anarchy'' - the stationary distribution of
the corresponding aggregate demand process has a smaller tail. This tail, which corresponds
to rare and undesirable demand spikes, is important in many applications of interest.
%%
% On the other hand, when the agents can cooperate with each other in optimizing their
% total costs, a higher market efficiency is achieved at the cost of a higher probability
% of demand spikes.
%

With a better understanding of the efficiency-risk tradeoff, we investigate, in a
non-cooperative setup, how resource pricing can be used as a tool by the system operator
to tradeoff between efficiency and risk.

We further provide a convex characterization of the Pareto front of different system
performance measures. The Pareto front determines the tradeoff among volatility
suppression of concerned measurements in the system with load scheduling dynamics. This
is the fundamental tradeoff in the sense that system performance achieved by any load
scheduling strategies induced by any specific market architectures is bounded by this
Pareto front.

% We thus posit that the origins of endogenous risks in such systems may lie in the market
% architecture, which is an inherent characteristic of the system ; and a specific market
% architecture leads to the agent load scheduling scheme, which determines the system
% performance measures.

%%
% When the system operator is able to differentiate consumers and charge different prices
% for individuals, ``congestion pricing'' can be used to affect the agent behavior so
% that the system outcome is more like that in the cooperative market, namely higher
% efficiency and higher cost.  % When the system operator is only able to charge a single
% price for all consumers, we formulate the problem for the system operator and show
% numerically the Pareto front of all linear pricing schemes.

%\end{abstract} 

\end{abstractpage}

% Additional copy: start a new page, and reset the page number.  This way,
% the second copy of the abstract is not counted as separate pages.
% Uncomment the next 6 lines if you need two copies of the abstract
% page.
% \setcounter{page}{\thesavepage}
% \begin{abstractpage}
% \input{abstract}
% \end{abstractpage}

\newpage

\section*{Acknowledgments}

The warmest thanks go to my advisor Munther Dahleh and Mardavij Roozbehani for their
guidance, understanding, patience, and friendship during my first two years at MIT.
They always encouraged me to ask meaningful and interesting questions, to think deeply
and to think out of the box, and the great experience of working with them motivates me
to move on to my PHD study.

I would also like to thank LIDS, where the great learning environment provides positive
feedback to one's insatiable curiosity, and talking to those smart and nice people can
make one's worrieness and stress disappear.

Finally, and most importantly, I would like to thank my wonderful parents, for their
faith in me, and for their unending love and support.

% % Some departments (e.g. 5) require an additional signature page.  See
% % signature.tex for more information and uncomment the following line if
% % applicable.
% % \include{signature}
 \pagestyle{plain}
   % -*- Mode:TeX -*-
%% This file simply contains the commands that actually generate the table of
%% contents and lists of figures and tables.  You can omit any or all of
%% these files by simply taking out the appropriate command.  For more
%% information on these files, see appendix C.3.3 of the LaTeX manual. 
\tableofcontents
\newpage
\listoffigures
\newpage
\listoftables

 %% sec: intro

% \tc{red}{Or maybe we start as follows for the motivation part:
% \\
% 1. importance of examining the endogenous risk in MAS
% \\
% 2. introduce our stylized model of price coupling
% \\
% 3. point out the possible applications of the stylized model
% \\}

%\section{Introduction}
\chapter{Introduction}
\label{sec:introduction}

Load scheduling, i.e., optimizing the demand for a resource over multiple periods to
minimize the expected total cost of consumption, plays a crucial role in a wide array of
applications, including dynamic demand response to realtime prices in electricity markets
\cite{cohen1988optimization, roozbehaniintertemporal}, load scheduling in cloud computing
under QoS constraints \cite{buyya2008market, li2009utility, buyya2002deadline}, and
multi-period rebalancing of multiple portfolio accounts in the presence of transaction costs
\cite{stubbs2009multi}.
In many cases where the price per unit resource in each period is determined by the
instantaneous aggregate demand of finitely many agents, the problem falls into the category
of dynamic oligopolistic competition \cite{maskin1987theory, leontief1936stackelberg}.

%{\tiny setup briefly, define the tail risk that we look at.}

In a multi-agent system, profit-seeking agents try to maximize their own utilities, by
forming rational expectations over the behaviors of other agents, and responding to
instantaneous changes in the environment.
The agent load scheduling scheme at equilibrium is shaped by different features of the
oligopolistic market architecture, including whether the agents are able to cooperate in
decision making, including the risk sensitivity of the agents, and including how their
costs are coupled, namely, the rule that the price is determined.
%
% From an individual agent's perspective, scheduling loads in response to instantaneous prices
% has economic value \cite{roozbehaniintertemporal}.
From a system operator's perspective, the impact of the aggregate behavior of
rational agents
% , who aim at maximizing their own utilities through optimized dynamic
% responses to the changes in the environment,
%leveraging the uncertainties in the environment,
is nontrivial -- on one hand, it determines the system efficiency, and on the other hand,
agent interactions can lead to endogenous risk. For example, in electricity markets,
aggregate demand spikes can incur additional costs to the resource provider or the power
system as a whole.
% ; in financial markets, a price surge due to large
% trading quantities can cast a negative impact on stock market stability.
%\tc{red}{add reference here}
%%
We shall focus on the measure of risk that quantifies such aggregate demand spikes, and
examine how they may arise from the market architectural properties.

%{\tiny literature on endogenous risk}

In many complex systems with interactive agents, for example, power networks, financial
markets, social networks, and biological networks, the mechanisms that can possibly channel
exogenous shocks into endogenous risk are still not well understood. Previous research
efforts have explored various possible origins of endogenous risk.
The notion of ``endogenous risk'' in financial market was introduced in
\cite{danielsson2003endogenous, danielsson2011endogenous}. When homogeneous traders with
trading limits start to sell as the price decreases, their failure to endogenize other
traders' actions leads to price fluctuation and instability. The authors argue that ignoring
the feedback link from traders' actions to the market price can damage the financial market
in this way. Other research efforts that attempted to explain the fluctuations in financial
market have examined information asymmetry \cite{chae2005trading}, bounded rationality
\cite{nottola1994dynamics} and heterogeneous beliefs \cite{geanakoplos2009leverage}.
In our work, we assume rational agents, who are fully aware of the pricing mechanism,
have complete information about other agents in the market, and form rational
expectations.  In this work, we provide an alternative explanation through a comparative
study, and posit that endogenous risks can arise from the nature of the system dynamics
even at a complete information rational expectation equilibrium (REE).
%%

%\tc{red}{to here}

% \tc{red}
%{\tiny what we did}

We create an abstract dynamic framework to model agents' response to realtime costs in
the form of load scheduling with deadline constraints, and we investigate the impact of
aggregate behavior on system performance, with the hope of finding behaviors and
properties that transcend the abstraction of the model.
We first examine the system performance under the non-cooperative and cooperative market
architectures, both with marginal production cost pricing of the resource so that agents'
demands for the resource are strategic substitutes.
Under the non-cooperative market architecture, the load scheduling problem is formulated as
a stochastic dynamic oligopolistic game, and under the cooperative market architecture, it
is formulated as an infinite-horizon average-cost Markov decision problem (MDP).
We shall focus on two performance measures: market efficiency and the risk of aggregate
demand spikes.
In the non-cooperative market, each agent schedules his consumption to optimize his expected
cost of implementing his schedule; in the cooperative market, the agents cooperate in the
decision making process to optimize aggregate expected cost.
We observe that under the cooperative market architecture, the agents are more aggressive in
absorbing exogenous uncertainties, and they can achieve higher market efficiency, i.e.,
lower cost on average. However, the tradeoff is a higher endogenous risk in terms of a
higher probability of aggregate demand spikes.
We also show that across load scheduling strategies induced by various oligopolistic market
architectures, there exists a tradeoff between efficiency and risk.

With a better understanding of the origin of the aggregate demand spikes, we facilitate
the analysis by focusing on the linear time-invariant part of the system dynamics and
defining the substitute performance measures. In the linear time-invariant framework, we
examine how the pricing rule can be used to induce the desired agent behavior in a
non-cooperative market. Moreover, we characterize the Pareto front of system performance
measures, which describes the fundamental tradeoff limit for the system with the load
scheduling dynamics.
%
% Moreover, we show that when the system operator is able to differentiate consumers and
% charge different prices for individuals, ``congestion pricing'' can be used to affect the
% agent behavior so that the system outcome is more like that in the cooperative market.
% %
% When the system operator is only able to charge a single price for all consumers, we
% formulate the problem for the system operator and show numerically the Pareto front of
% all linear pricing schemes.
% %%
The implication of our efficiency and risk analysis is that when the system architecture
and operational policies are designed, system efficiency should not be the only goal that
is pursued; endogenous risk and the associated tradeoffs should also be carefully
considered.
%

%\tc{red}
%{\tiny find applications}

An interesting example where we can apply the analytical framework to study the
efficiency-risk tradeoffs is the dynamic demand response to realtime prices in
electricity markets in the form of scheduling flexible loads.
% The advent of distributed renewable energy sources poses new challenges for the design and
% operation of the future smart grids.
On the supply side, the intermittency of the renewable sources introduces exogenous supply
shocks.  On the demand side, large or perhaps small consumers may be able to actively
respond to the realtime eletricity prices. A considerable amount of the consumer response
will take the form of scheduling flexible loads, for example, electrical vehicle charging,
building heating, and industrial processing \cite{belhomme2008address, kim2011scheduling,
mohsenian2010optimal}.
%%
% The uncertainties lie in the renewable generations, the load requirements of different
% agents, and the timing of consumers' entrance and departure from the market; and the
% system dynamics comes from the individual's load scheudling over multi-periods.
A specific example of electrical vehicle charging where our framework fits can be found in
\cite{foster2010energy}.
% Setups similar to our work can be found in \cite{foster2010energy, caramanispower,
% jiang2011multi}.
%
%\tc{blue}
% Assuming inelastic demand, the strategic bidding behavior of suppliers to withhold
%   generation capacity and drive up prices was studied in \cite{berry1999analyzing,
%   guan2001gaming}.  However, we
We model the market participation behavior of both the consumers and the distributed
renewable generations, with potential load scheduling and storage techniques. The
resulting dynamic demand supply interaction can better model future smart grids.
Consumer participation in smart grids is modeled in a similar way in
\cite{couillet2012mean}, but the heterogeneous deadline constraints of individual
players, which are essential in producing the aggregate demand spikes in our framework,
are not modeled explicitly there.
However, this is important, as in electricity markets, exceedingly large demand and/or
price spikes introduce a level of volatility that can not only cause serious economic
damage to both the reliable service provider and consumers, but also undermine viability
of power markets as a whole.

The remainder of the thesis unfolds as follows.
In Chapter \ref{sec:system-model}, we introduce the system model and formulate the
problem;
in Chapter \ref{sec:example-l=2}, we focus on a specific case for which analytical
solutions are obtained, and examine how various architectural properties affect the
efficiency-risk tradeoffs;
in Chapter \ref{sec:pricing}, we introduce the linear time-invariant framework, and
discuss how the system operator's decision on the pricing rule will affect agent load
scheduling behavior in a non-cooperative setup;
in Chapter \ref{sec:generalL}, we provide a convex characterization of the Pareto front
of performance measures, which dictates the fundamental tradeoff of the system with load
scheduling dynamics;
in Chapter \ref{sec:conclusion}, we conclude the paper with a discussion about future
work.

\chapter{System Model}
\label{sec:system-model}

% In this section
In this chapter
we introduce the general system model consisting of heterogeneous agents
which arrive at the system following a random arrival process, a reliable resource
provider and a marginal cost pricing mechanism. We also define the non-cooperative and
cooperative market architectures.

%\subsection{Agent Arrival Process}
\section{Agent Arrival Process}
\label{sec:agent-arriv-proc}
We analyze a market model in which the agent arrival process is a discrete time random
process with time intervals indexed by $t=0,1,2,\cdots$. When an agent arrives, he
activates a job that requires consuming a certain amount of the resource to complete. The
agent has to finish the job within a finite window of time, and leave the market at his
deadline.
We define the number of periods that an agent stays in the market to be his {\sl type},
denoted by $l\in\mathcal L = \{1,\dots,L\}$.  We assume that agents of type $l$ arrive
according to a Bernoulli process $\{h_l(t):t\in\mbb Z\}$, with rate $q_l$.
Upon arrival at the beginning of period $t$, an agent carries a job which requires
$d_l(t)$ units of the resource in total. We assume that the sequence $\{d_l(t):t\in\mbb
Z\}$ is i.i.d., drawn from a general distribution $ D_l$ with mean $\mu_l = \mathbb
E[D_l]$, variance $\sigma_l^2 = \text{Var}[D_l]$, and with support over the set of all
real numbers $\mbb R$.
Let the $L$-dimensional column vectors $\mb h(t) = [h_l(t)]\in\{0,1\}^{L}$, and $\mb d(t) =
[d_l(t)]\in\mbb R^{L}$ denote the vector forms of arrival events and the corresponding
workloads. Let $U(t)$ denote the instantaneous aggregate demand for the resource from all
agents in the market.
The key notations that we will introduce throughout the paper are listed in Table
\ref{table:notations}.

\begin{Rem}
  Note that for the convenience of our analysis, we allow the load realizations as well
  as the instantaneous resource demand from the agents to become negative. This models
  the situation where distributed agents can be both suppliers and consumers in the
  market. 
  In financial market, the informed traders can be both buyers and sellers, and the
  uninformed traders have a passive role which is similar to the reliable resource
  provider \cite{grossman1980impossibility}.
  In electricity markets, this corresponds to the scenario where consumers are equipped
  with distributed renewable generations or pumped-storage units, and are able to sell
  energy back to the power grid.
  We ran extensive numerical simulations for the scenario where there is a lower bound on
  instantaneous resource demand and/or supply. In particular, when the lower bound equals
  zero, the agents are only consumers and cannot supply the resource to the market.
  In all of our the simulations, the main results hold qualitatively.
\end{Rem}

%\subsection{Resource pricing}
\section{Resource pricing}
We assume that there is a reliable resource provider which always produces enough amount
of the resource to meet the aggregate demand in each period. Moreover, we assume that the
production cost borne by the provider is of quadratic form $\frac{1}{2}U(t)^2$, and the
price per unit resource, $p(t)$, is set to be the marginal cost of production in each
period, thus $p(t)=U(t)$.
We adopt quadratic cost functions for two reasons: firstly they constitute second-order
approximation to other types of nonlinear cost functions, and secondly they are
analytically tractable, with which closed-form solutions can hopefully provide insights
into more general system dynamics.
% Note that we can also interpret the proportional pricing as a result of linearizing a
% production cost function in general form around the steady state production level, with
% the coefficients determined by the first order derivative of the cost function
% evaluated at the steady state.
%
Also, note that the quadratic cost function only models the production cost of the
reliable resource provider, which we assume to have no intertemporal
constraints. Overall, the aggregate demand is satisfied by the sum of distributed
supplies from the agents, and the resource produced by the reliable resource
provider. The price is set to provide sufficient incentive to the reliable resource
provider to produce at the level where the overall production matches the aggregate
consumption.
In electricity markets, marginal cost pricing is a widely used mechanism
\cite{schweppe1988spot}. When both the suppliers and consumers are price takers and there
is no intertemporal ramping cost, marginal cost leads to social optimality.
Moreover, the reliable resource provider corresponds to the conventional electricity
generations which provide reliable electricity, as opposed to the distributed renewable
generations, which are stochastic in the nature.

%\subsection{System State Evolution}
\section{System State Evolution}
At any period $t$, we group the agents by their departure times.  For any
$\tau\in\mathcal L$, there are at most $(L+1-\tau)$ agents who will stay in the market
for $\tau$ periods (including $t$). They correpond to the type $\tau$ arrival at time
$t$, the type $(\tau+1)$ arrival at time $(t-1)$, etc.
Take $L=5$, $\tau=3$ as an example. Figure~\ref{fig:D1} shows that at time $t$ there are
3 possible agents who will stay in the market for $\tau=3$ periods.
For notational convenience, we index a type $l$ agent who at time $t$ will continue to
stay in the market for $\tau$ periods by a tuple $(l,\tau)_t$, and we list all possible
$(l,\tau)$ tuple in the ordered set: 
\begin{align*}
  \mathcal C = \{(1,1), (2,1),(3,1)\cdots,&(L,1),
  \\
  (2,2), (3,2),\cdots,& (L,2),
  \\
  \cdots,&(L,L). \}
\end{align*}
Let $D_c = L(L+1)/2$ denote the cardinality of the ordered set $\mathcal C$.  Let
$u_{(l,\tau)}(t)\in\mbb R$ denote the \textsl{instantaneous demand} from agent $(l,
\tau)_t$, with the vector form denoted by: 
\[\mb u(t) =[u_{(l, \tau)}(t): (l,\tau)\in\mathcal C]\in\mbb R^{D_c}.\]
If at time $t$ there is no agent $(l,\tau)_t$, i.e., $h_l(t+\tau-l)=0$, we simply define
$u_{(l,\tau)}(t) = 0 $. The instantaneous aggregate demand is therefore $U(t) =
\sum_{(l,\tau)\in \mc C}u_{(l, \tau)}(t) = \mb 1'\mb u(t)$, where $\mb 1$ is a
$D_c$-dimensional column vector of all ones.
Similarly, we define the \textsl{backlog state} $\mb x(t)$ and the \textsl{existence
  state} $\mb o(t)$ as follows:
\begin{align}
  &\mb x(t) = [x_{(l, \tau)}(t): (l,\tau)\in\mathcal C]\in\mbb R^{D_c},
  \label{eq:def-x}\\
  &\mb o(t) = [o_{(l, \tau)}(t): (l,\tau)\in\mathcal C]\in \{0,1\}^{D_c},
  \label{eq:def-o}
\end{align}
where element $x_{(l, \tau)}(t)$ denotes agent $(l, \tau)_t$'s unsatisfied load at time
$t$, and  element $o_{(l, \tau)}(t) = 1$ if and only if there is an arrival of type $l$
agent at time $(t+\tau - l)$.
Finally, system state at time $t$ is defined to be $ \mb s(t) = (\mb x(t), \mb o(t))\in
\mathcal S$, where $\mathcal S =\mbb R^{D_c }\times\{0,1\}^{D_c} $ is the state space.
% Similarly, we define the \textsl{backlog state} to be a vector:
% \[\mb x(t) = [x_{(l, \tau)}(t): (l,\tau)\in\mathcal C]\in\mbb R^{D_c},\]
% with element $x_{(l, \tau)}(t)$ denoting agent $(l, \tau)_t$'s unsatisfied load at time
% $t$.  We also define the \textsl{existence state} to be:
% \[\mb o(t) = [o_{(l, \tau)}(t): (l,\tau)\in\mathcal C]\in \{0,1\}^{D_c}, \vspace{-3pt}
% \]
% with element $o_{(l, \tau)}(t) = 1$ if and only if there is an arrival of type $l$ agent at
% time $(t+\tau - l)$.  Finally, system state at time $t$ is defined to be $ \mb s(t) = (\mb
% x(t), \mb o(t))\in \mathcal S$, where $\mathcal S =\mbb R^{D_c }\times\{0,1\}^{D_c} $ denotes
% the state space.
%
We assume that system state is updated after the realization of $\mb h(t)$ and $\mb d(t)$
at the beginning of each period $t$, and the state information is publicly available to
all agents in the market\footnote{ We acknowledge that this complete information
  assumption is very strong in real life applications with autonomous agents, especially
  when the number of agents is large.  Information structure, though an important issue
  in dynamic games, is not the focus of this paper, as the identified mechanism that
  produces endogenous risk of spikes also exists in incomplete information models. This
  simplification assumption affords us a model which is tractable and can serve as a
  benchmark for incomplete information models.}. The system state $\mb s(t) = (\mb x(t),
\mb z(t))$ evolves as follows:
\begin{align}
  \mb x(t+1 ) &= \mb R_1(\mb x(t) - \mb u(t)) + \mb R_2\mb d(t)
    \label{eq:state-evolve-gen1}
    \\
    \mb o(t+1) &= \mb R_1\mb o(t) + \mb R_2\mb h(t)
    \label{eq:state-evolve-gen2}
\end{align}
where $\mb R_1$ is a  $D_c\times D_c$ matrix with non-zero elements:
\begin{align*}
  &
  \mb R_1\Big((k-1)(L+\frac{2-k}{2})+i+1,\ \ \ k(L+\frac{1-k}{2})+i \Big) = 1,
\\
&\qquad\qquad\qquad \text{for all } 1\le i\le L-k \text{ and } 1\le k\le L-1,
\end{align*}
and all other elements being $0$.  Also, $\mb R_2$ is a $D_c \times L$ matrix with
non-zero elements:
\begin{align*}
%  \small{\text{$
  \mb R_2\Big((l-1)(L+\frac{2-l}{2}) +1, \ \ \ l \Big) = 1, \ \text{for all } 1\le l\le L .
  %$}}
\end{align*}
and all other elements being $0$. 
As an example, the matrices $\mb R_1$ and $\mb R_2$ for $L=3$ are given in
Appendix~\ref{sec:example-state-space-L3}.
%where $\mb R_1$ is a constant $D_c\times D_c$ matrix defined as follows:
% \begin{align*}
%   &\mb R_1 =  
%   %\\ &
%   \left[
%     \begin{array}{llllll}
%       \mb 0_{1\times D_c}  &&&   
%       \\ 
%       \mb 0_{(L-1)\times L}  &    \mb I_{(L-1)\times (L-1)} & & \mb 0_{(L-1)\times (D_c-L-(L-1))} 
%       \\ 
%       \mb 0_{1\times D_c}  &&&    
%       \\
%       \mb 0_{(L-2)\times (L + L-1)}  & &  \mb I_{(L-2)\times(L-2)} &  \mb 0_{(L-2)\times (D_c-L-(L-1)-(L-2))}
%       \\
%       \mb 0_{1\times D_c}  &&&   
%       \\
%       &    &      \ddots           &                  
%       \\
%       \mb 0_{1\times(L + L-1 +\cdots + 2)}  &  &        & \mb I_{1\times 1}                 
%       \\
%       \mb 0_{1\times D_c}  &&&   
%     \end{array}
%   \right]
% \end{align*}

%\subsection{Non-cooperative Market Architecture}
\section{Non-cooperative Market Architecture}
We define the {\sl non-cooperative market architecture} to be a market setup in which
there is no coordination among the strategic agents in scheduling their loads. With full
information about the system model and the state evolution $\{\mb s(t'): t'\le t\}$,
an agent $(l,\tau)_t$ makes the decision of his instantaneous resource demand
$u_{(l,\tau)}(t)$ based on his observation of system state $\mb s(t)$.
% Similar to the model in \cite{foster2010energy},
We assume that the agents do not directly derive utility from consumption of the
resource. Thus the only objective they have is to minimize the expected total cost, under
the constraint that each agent's total consumption by his deadline must be equal to his
workload.
Note that compared to the standard modeling of utility as an increasing function in
consumption, this is a more accurate modeling of consumer behavior in terms of decision
making about electricity consumption. 
Our framework can also be extended to cases where agents value their consumptions. For
example, later in Chapter~\ref{sec:generalL}, we shall relax the deadline constraints,
while including the disutility from the mismatch between real consumptions and the target
consumption to complete the tasks into agent payoff function.

More specifically, under the non-cooperative architecture, a type $l$ agent who arrives
at time $t$ dynamically optimizes his consumption schedule $\{u_{(l,l-i)}(t+i):\
i=0,1,\dots,l-1\}$ to minimize his expected payment $\mathbb
E[\sum_{i=0}^{l-1}p(t+i)u_{(l,l-i)}(t+i)]$.
% by solving an optimization problem:
% \begin{align*}
%   \min_{\left\{
%       %\begin{array}{l}
%         u_{(l,l-i)}(t+i):\ =0,1,\dots,l-1   
%       %\end{array}
%     \right\}} \mathbb E & \left[\sum_{i=0}^{l-1}p(t+i)u_{(l,l-i)}(t+i)\right]
%   \\
%   \text{subject to: }& \sum_{i=0}^{l-1}u_{(l,l-i)}(t+i) = d_{l}(t),
%   \\
%   & p(t) = \sum_{(l,\tau)\in\mathcal C}u_{(l,\tau)}(t),\quad \fa t
% \end{align*}
% where the price $p(t)$ is determined by the aggregate demand for the resource in period
% $t$: $p(t) = U(t)$.
%%
%\tc{blue}{
% The non-cooperative market architecture leads to a\
Due to the cost coupling through endogenous pricing, we model agent interaction by a
stochastic dynamic game, with the following specificiation:
  \begin{itemize}
  \item \tb{Players:} Over infinite time horizon, the players are indexed by $\{(l,\tau)_{t}:
    t\in \mbb Z, (l,\tau)\in\mathcal C\}$ according to their type and arrival time in the
    market.    
  \item \tb{State Space:} The state space is given by $\mathcal S$.
  \item \tb{Action Set:} The action set is given by $\mathcal A$. In particular, the action set
    of player $(l, \tau)_t$ at time $t$ in state $\mb s$ is given by:
    % http://en.wikipedia.org/wiki/Extensive-form_game
    \begin{align}
      A_{(l,\tau)}(\mb s) = \left\{
        \begin{array}[l]{ll}
          0, & \text{if } z_{(l,\tau)} = 0 
          \\
          x_{(l,\tau)}, & \text{if } z_{(l,\tau)}=1 \text{ and } \tau=1
          \\
          \mbb R, & \text{otherwise}
        \end{array}
      \right.
      \label{eq:action_set}
    \end{align}
  \item \tb{Transition Probability:} For each state $\mb s$ and action vector $\mb
    u\in\prod_{(l,\tau)} A_{(l,\tau)}(\mb s)$, the transition probability $\mbb P(\mb s'|\mb s,
    \mb u)$ is consistent with the state dynamics in (\ref{eq:state-evolve-gen1}),
    (\ref{eq:state-evolve-gen2}) and the agent arrival process in \ref{sec:agent-arriv-proc}.
    % \[
    %  P(\mb s' | \mb s, \mb u) =
    % \int
    % \mb 1[\mb x' = \mb R_1(\mb x-\mb u)+\mb R_2\mb d]
    % d P(\mb d)
    % \cdot
    % \int
    % \mb 1[\mb o' = \mb R_1 \mb o + \mb R_2\mb h]
    % d P(\mb h)
    % \]
  \end{itemize}
  We shall focus on {\sl Markov Perfect Equilibrium} (MPE) \cite{shapley1953stochastic,
    maskin1988theory} throughout our discussion. This refers to a subgame perfect
  equilibrium of the stochastic dynamic game where players' strategies only depend on the
  current state.  The {\sl Markov strategy} is thus defined as a function:
  \[\mathbf u%_{(l,\tau)_t}
  :\mathcal S\to \mathcal A\]
  which maps the system state to the instantaneous demand in the action set from agent
  $(l,\tau)_t$.
  Moreover, as all agents have the same cost structure, it is natural to focus on
  symmetric stationary pure strategy equilibria where for every $(l,\tau)\in\mathcal C$,
  the agents $\{(l,\tau)_t:t\in\mbb Z\}$ adopt the same decision rule denoted by $\mb
  u(\mb s)$.
  The symmetry of this problem makes it possible to consider a single agent's problem to
  characterize the equilibrium, which we formalize as follows:
%
%
% We formally define the Markov Perfect Symmetric Equilibrium Strategy in our problem as follows:
%}
\begin{Def}[Markov Perfect Symmetric Equilibrium Strategy]
\label{def:mpe}  
A strategy profile
\[\mb u^{nc} = \{u_{(l,\tau)}^{nc}(\mb s): (l,\tau)\in\mathcal C, \mb s\in\mathcal S\}\]
is defined to be a Markov Perfect Symmetric Equilibrium Strategy, if the following fixed
point equations are satisfied for all agents $(l,\tau)\in\mathcal C$ at any time $t$, for
any system states $\mb s(t)\in\mathcal S$:
\begin{align}
  u_{(l,\tau)}^{nc}(\mb s(t)) = \arg\min_u \mathbb E\Big[ p(t) u + \sum_{i = 1}^{\tau -1}
  p(t + i) u_{(l,\tau-i)}^{nc}(\mb s(t + i))  \Big| \mb s(t) \Big]
  \label{eq:mpe-fix-point}
\end{align}
\vspace{-30pt}
\begin{align}
  \text{subject to: } &\sum_{i=0}^{l-1}u_{(l,l-i)}^{nc}(\mb s(t+i)) = d_{l}(t), \quad \fa t,l,
  \nonumber\\
  & p(t) = {u + \hspace{-5pt}\sum_{(l',\tau')\in \mc C, (l',\tau')\neq (l,\tau)} \hspace{-10pt}
    u_{(l',\tau')}^{nc}(\mb s(t))},
  \nonumber\\
  & p(t+i) = \sum_{(l',\tau')\in \mc C} u_{(l',\tau')}^{nc}(\mb s(t+i)) , \quad \fa i\ge 1, \nonumber
\end{align}
where $\mb s(t)$ evolves according to (\ref{eq:state-evolve-gen1}),
(\ref{eq:state-evolve-gen2}).
% \begin{align}
%     u_{(l,\tau)}^{nc}(\mb s(t)) = \arg\min_u &\mathbb E\Big[ u \Big(\underbrace{u+
%       \sum_{(l',\tau')\neq (l,\tau)}u_{(l',\tau')}^{nc}(\mb s(t))}_{p(t)}\Big)
%     \nonumber
%     \\    
%     + \sum_{i =
%       1}^{\tau -1} \Big( u_{(l,\tau-i)}^{nc}&(\mb s(t+i)) \underbrace{\sum_{(l',\tau')}
%       u_{(l',\tau')}^{nc}(\mb s(t+i)) }_{p(t+i)} \Big) \Big| \mb s(t) \Big]
%       \label{eq:mpe-fix-point}\\
%       \text{subject to: } &\sum_{i=0}^{l-1}u_{(l,l-i)}^{nc}(\mb s(t+i)) = d_{l}(t), \quad \fa t,l,
%       \nonumber
%       \\
%       \text{where } \mb s(t)& \text{ evolves according to
%          (\ref{eq:state-evolve-gen1}), (\ref{eq:state-evolve-gen2})}.
%     \nonumber
%   \end{align}
\end{Def}

% \begin{Prop}[Existence of equilibrium strategy]
%   \label{prop:exist-equil-general}
%   % Assume that the price $p(t)$ is proportional to the instantaneous aggregate demand
%   % $U(t)$,
%   % and that agents of each type arrive in the market following independent Bernoulli
%   % processes
%   % with rate $q_l$, and workloads of each type are i.i.d. distributed according to
%   % distributions
%   % $\mathcal D_l$,
%   % \tc{blue}{
%     Under the assumptions of marginal cost pricing, agent arrival process and workload
%   distributions, there exists a Markov perfect symmetric equilibrium strategy $\mb u^{s}(\mb
%   s)$ which satisfies the fixed point equations in (\ref{eq:mpe-fix-point}).
% %}
% \end{Prop}

% \begin{Proof}
%   % convexity of the the cost function, and job deadline, leads to stationary distribution.
%   Appendix~\ref{prf:exist-equil-general}
% \end{Proof}

% \vspace{5pt} The equilibrium strategy $\mb u^{nc}(\cdot)$ can be found via numerical
% iteration techniques, and the computational complexity is exponential in $D_c$.

%\subsection{Cooperative Market Architecture}
\section{Cooperative Market Architecture}
As an efficiency benchmark, we consider the cooperative market architecture, under which the
agents can coordinate their actions to minimize their aggregate expected cost. Later, we show
that under the assumptions of quadratic production cost and marginal cost pricing, the
cooperative market architecture leads to the highest market efficiency,
%\tc{blue}{
defined as the total surplus from all agents and the reliable resource provider.
% }
The cooperative market architecture can model the scenario where the agents agree a
priori upon a common strategy that minimizes their aggregate expected cost, and respond
to the realtime market conditions according to the prespecified strategy. %
Particularly, in future electricity markets, the cooperative scheme may correspond to the
situation where the consumers with flexible loads pass all the relevant information to a
load aggregator who schedules the loads on their behalf. We are interested in the system
performance in the stationary equilibrium, and define the {\sl optimal stationary
  cooperative strategy} under the cooperative market architecture as follows:
\begin{Def}[Optimal Stationary Cooperative Strategy]
  \label{def:opt_coop}
  A strategy profile
  \begin{align*}
    % & \mb u^c =\{\mb u^c(\mb s): \mb s\in\mathcal S\}
    & \mb u^c =\{\mb u^c_{(l,\tau)})(\mb s):(l,\tau)\in\mathcal C, \mb s\in\mathcal S \}
  \end{align*}
  is defined to be an Optimal Stationary Cooperative Strategy if $\mb u^c(\mb s) =
  \big[u_{(l,\tau)}^c(\mb s): (l,\tau)\in\mathcal C\big]$ solves the following fixed point
  equations for any system states $\mb s(t)\in\mathcal S$:
  %\tc{blue}{
  \begin{align}
    % \{u_{(l,\tau)}^c(\mb s(t)): (l,\tau)\in\mathcal C\} =&
    \mb u^c(\mb s(t)) =&
    \arg\min_{\mb u^c =\big[ u_{(l,\tau)}: (l,\tau)\in\mathcal C \big]}
    \lim_{T\to\infty}\frac{1}{T-t} \mathbb
    E\Big[
    %\nonumber\\ &
    \sum_{(l,\tau)\in \mc C}p(t) u_{(l,\tau)} +
%    \nonumber\\
    \sum_{t'=t+1}^{T} \sum_{(l,\tau)\in \mc C} p(t+i) u_{(l,\tau)}^c(\mb s(t')) 
    \Big| \mb s(t)
    \Big]
    \label{eq:CSL-gen}
    \\ \text{subject to: }
    & \sum_{i=0}^{l-1}u_{(l,l-i)}^c(\mb s(t+i)) = d_{l}(t), \quad \fa t,
    l, \nonumber\\
    & p(t) = \sum_{(l, \tau)\in \mc C}u_{(l,\tau)}, \nonumber\\
    & p(t+i) = \sum_{(l, \tau)\in \mc C}u_{(l,\tau)}^c(\mb s(t+i)),\ \ \fa i\ge 1, \nonumber
  \end{align}
  where $\mb s(t)$ evolves according to (\ref{eq:state-evolve-gen1}),
  (\ref{eq:state-evolve-gen2}). 
 % }
\end{Def}
The above problem is a standard infinite horizon average cost MDP, and the associated
Bellman equation can be solved via standard value iteration or policy iteration
\cite{bertsekas2011dynamic}.
%\tc{red}{should I write down the bellman here or not?}

%\subsection{Welfare Metrics}
\section{Welfare Metrics}
\label{sec:welfare-metrics}
Different oligopolistic market architectures induce different agent behaviors, which lead
to different stationary distributions of the aggregate demand process $\{U(t):t\in\mbb
Z\}$. We shall focus on two welfare metrics: \tb{efficiency} and \tb{risk}. More
specifically, we define {\sl efficiency} to be the expected sum of the resource
provider's surplus $W_p$ and the agents' surplus $W_a$ as follows:
% \begin{align}
%   W& = \underbrace{\mbb E[ p(t)U(t) - \frac{1}{2}U(t)^2]}_{W_p} + \underbrace{\mbb
%     E[-p(t)U(t)]}_{W_a}
%   \nonumber\\
%   & = -\frac{1}{2}\mbb E[U(t)^2]
%   \nonumber\\
%   & = \frac{1}{2} W_a
% %  & = -\frac{1}{2}\left(\text{Var}[U(t)] + (\sum_{l}q_l\mu_l)^2\right)
%   \label{eq:def-efficiency}
% \end{align}
\begin{align}
  W = \underbrace{\mbb E[ p(t)U(t) - \frac{1}{2}U(t)^2]}_{W_p} + \underbrace{\mbb
    E[-p(t)U(t)]}_{W_a}
   = -\frac{1}{2}\mbb E[U(t)^2]
   = \frac{1}{2} W_a
  \label{eq:def-efficiency}
\end{align}
Note that under the assumptions of quadratic production cost and marginal cost pricing,
efficiency is decreasing in $\mbb E[U(t)^2]$.
%the second moment of the aggregate demand process.
%
In (\ref{eq:CSL-gen}), the optimal stationary cooperative strategy $\mb u^c(\cdot)$
maximizes $W_a$, thus achieves the highest efficiency in the sense of
(\ref{eq:def-efficiency}), which we denote by $W^c = W_p^c + W_a^c$. Let $W^{nc} =
W_p^{nc} + W_a^{nc}$ denote the efficiency achieved by the equilibrium strategy $\mb
u^{nc}(\cdot)$ under the non-cooperative market architecture. Note that $W^{nc}\le W^c$
and $W_a^{nc}\le W_a^{c}$.
This efficiency loss $W_a^{c}-W_a^{nc}$ is commonly known as the ``price of anarchy'' due
to the strategic behavior of non-cooperative agents when payoff externalities exist.

We define {\sl risk} to be the tail probability of the stationary process of aggregate
demand:
\begin{align}
  R = \Pr(U(t)>M)
  \label{eq:def-risk}
\end{align}
for some positive large constant $M$. As a result of marginal cost pricing and increasing
marginal cost, risk also captures the tendency for aggregate demand / prices to spike
drastically (above a large $M$).  We also define {\sl market robustness} to be:
\begin{align}
  B = 1-R. 
  \label{eq:robust-def}
\end{align} 
Apart from market efficiency, risk, in terms of demand spikes, is also an important
welfare metric.  %
In a given oligopolistic market, rational agents respond to endogenous realtime prices to
minimize individual costs.
% , without explicitly considering the externality they create for
% each other, and for the system as a whole.
%
However, a system designer may have interests different from the agents, and be concerned
about the risk, in particular the aggregate demand spikes or cost surges.
In the sequel, we shall demonstrate, by analyzing the case with $L=2$, that under the
non-cooperative market architecture, even though there is a efficiency loss, the
strategic behavior also results in a smaller tail probability, which is associated with a
lower endogenous risk.
%%
%\tc{blue}{
A more fundamental question that we attempt to address is to what extent exogenous
uncertainties is inevitable and to what extent it can be controlled in the system. More
specifically, is there a limit of the feedback control, in the form of load scheduling,
to achieve the dual goals of increasing market efficiency and reducing endogenous risk?
Later we will show that for a broad class of load scheduling strategies, the exogenous
randomness cannot be completely eliminated, and the dual goals cannot be achieved
simultaneously.
  % \cite{lestas2010fundamental}
%}

So far, we have formulated the load scheduling problem as a stochastic dynamic
oligopolistic game under the non-cooperative market architecture, and as an infinite
horizon average cost MDP under the cooperative market architecture.
%
% In general, there are no closed form solutions to either of the two formulations, and
% numerical solutions involve exponential complexity. In the following section, we will
% look into the case where the number of types $L=2$, and the equilibrium strategy as well
% as the optimal cooperative strategy can be found explicitly .
In general, there are no closed form solutions to either of the two formulations, and
numerical solutions involve exponential complexity. In the following chapter, we will
look into the case where the number of types $L=2$, and the equilibrium strategy as well
as the optimal cooperative strategy can be found explicitly .

%\tc{red}{also add, in the modified system, we can obtain some analysis explicitly.}
 % \section{Tradeoff Analysis for $L=2$ Case}
\chapter{Tradeoff Analysis for $L=2$ Case}
\label{sec:example-l=2}

%\subsection{Equilibrium Strategy and Optimal Cooperative Strategy}
\section{Equilibrium Strategy and Optimal Cooperative Strategy}
When $L=2$, there are only two types of agents in the system: type 1 agents with
uncontrollable loads that must be satisfied upon arrival, and type 2 agents who have the
flexibility to split the consumption between two consecutive time periods. Under the
assumption of Bernoulli arrival process, at any time $t$, there are at most 3 agents in
the market, which are indexed as: $(1,1)_t$, $(2,1)_t$, and $(2,2)_t$. Among the three
agents, only the type 2 agent $(2,2)_t$ that arrives in the current period needs to make
a nontrivial decision, while the other two agents have no choice but to
empty their backlogs and leave the market.
%fulfill their backlogs and leave the market by the end of period $t$.

Note that this simple case still retains the two key features of the general
model. Firstly, since the active time window between any two consecutive type 2 agents
partially overlap, when a type 2 agent schedules his consumption, he needs to take into
account the action of the preceding type 2 agent, as well as to anticipate the reaction
of the succeding type 2 agent, in a similar way of the sequential Stackelberg competition
\cite{leontief1936stackelberg}; secondly, this dynamic system has exogenous uncertainties
in terms of agent arrivals and load realizations.
% , and the timing when the agents are active in the market. %
Considering the case of $L=2$ sheds light on understanding agent behaviors induced by
oligopolistic market architectures in the general setup. In electricity market, this case
with a few oligopolistic agents can be used to study the interaction among a few load
aggregators, each of which has considerable market power.
% \footnote{ For a discussion of the necessary and sufficient features of the system for
%   the tradeoff to exist, please see Section \ref{sec:intu-expl}.  }

We first simplify the notations. %
% Since the type 1 agents have no flexibility in scheduling the realized workload, we can
% incorporate the randomness of Bernoulli arrival of type 1 agents into the distribution
% $\mathcal D_1$ as a probability mass at 0.  \footnote{ However, this simplification
% does not apply to other types of agents with flexible loads, because even when the load
% realization $D_l$ is zero for some type $l>1$, the agent still has the freedom to
% produce and sell the resource in the market, as long as his net consumption is 0 by his
% deadline.  The difference is also exhibited in the specification of the action sets in
% (\ref{eq:action_set}).}  Without loss of generality, we set $q_1 = 1$, and denote the
% arrival rate of type 2 agents by $q = q_2$.
%
% We define the state of {\sl aggregate backlog} as:
% \begin{align*}
%   x(t) = x_{(1,1)} (t) + x_{(2,1)}(t), 
% \end{align*} 
%where without ambiguity, we denote $u_{(2,2)}(\mb s(t-1))$ by $u(t-1)$.
When agent $(2,2)_t$ schedules his consumption $(u_{(2,2)}(t), u_{(2, 1)}(t))$,
% he does not distinguish between agents $(1,1)_t$ and $(2,1)_t$, since the deadline
% constraints determine that $u_{(1,1)}(\mb s(t)) = x_{(1,1)}(t)$, and $u_{(2,1)}(\mb
% s(t))= x_{(2,1)}(t)$.  }
the sufficient statistics of system state for him is $(x(t),d_2(t))$, where $ x(t) =
x_{(1,1)} (t) + x_{(2,1)}(t) $ is defined as the {\sl aggregate backlog} state.
We also define a {\sl linear strategy} as a strategy profile $\mb u(\mb s)$ if
$u_{(1,1)}(\mb s)= x_{(1,1)}$, $u_{(2,1)}(\mb s)= x_{(2,1)}$, and $u_{(2,2)}(\mb s) =
u(x,d_2)$ which is a linear function of $x$ and $d_2$, i.e.,
\[u(x,d_2) = -ax + bd_2 + g.\]
% We will show that the Markov perfect symmetric equilibrium
% strategy and the optimal stationary cooperative strategy are both linear strategies.
%
% Finally, in each period $t$, the cost per unit consumption of the resource is given by $p(t) =
% U(t) =x(t) + u(t)$.

% For the  stochastic dynamic  oligopolistic game under the non-cooperative market
% architecture,
% For the non-cooperative case, the equilibrium strategy $u^{nc}(x,d_2)$ is characterized by
% the solution to the following fixed point equation:
% \begin{multline}
% u^{nc}( x(t), d_2(t)) = \arg\min_u\Big\{ u(u+x(t))
%   \\
%  + \mbb E\Big[(d_2(t)-u)  \Big( x(t+1) + h_2(t+1) u^{nc}(x(t+1),d_2(t+1))\Big)  
%   \Big| x(t), d_2(t) \Big] \Big\},
%   \label{eq:mpe2}
% \end{multline}
% where $x(t+1) = d_1(t+1)+(d_2(t)-u)$.

\begin{Prop}[Existence of linear MPE]
  \label{prop:exist-linear-mpe}
  For $L=2$, under the non-cooperative market architecture, there exists a Markov perfect
  symmetric equilibrium with the linear strategy $u^{nc}(x,d_2)$ given by:
\begin{align}
  u^{nc}(x,d_2) =&
  - \underbrace{\frac{1}{2(1+\sqrt{1-\frac{q_2}{2}})}}_{a^{nc}} x
  + \underbrace{\frac{1}{1+\frac{1}{\sqrt{1-\frac{q_2}{2}}}}}_{b^{nc}} d_2
  %\nonumber \\ &
  + \underbrace{\frac{q_1\mu_1 +
      q_2\mu_2\frac{1}{1+\sqrt{1-\frac{q}{2}}}}{2(1+\sqrt{1-\frac{q_2}{2}})}}_{g^{nc}}
  \label{eq:2-linear-equil}
\end{align}
\end{Prop}
\begin{Proof}
  Please refer to Appendix~\ref{prf:exist-linear-mpe}
\end{Proof}

% Next, we examine the cooperative scheme for the case $L=2$.
% By Definition \ref{def:opt_coop}, under the cooperative market architecture
The optimal stationary cooperative strategy can also be obtained as a closed form
solution of the Belllman equation with $L=2$.
% the following Belllman equation with value
% function $V^c(x)$ and average cost per period $\lambda^c$:
% \begin{align}
%   \lambda^c+V^c(x) =& (1-q)\Big(x^2 + \mathbb E_{d_1}\Big[V^c(d_1)\Big]\Big) +
%   q\mathbb E_{d_1, d_2}\Big[
%   %\nonumber\\ &
%   \min_{u}\big\{(x+u)^2 + V^c(d_2-u + d_1)\big\} \Big]
%   \label{eq:bellman2}
% \end{align}
\begin{Prop}[Existence of linear optimal stationary cooperative strategy]
  \label{prop:exist-coop-opt-L2}
  For $L=2$, under the cooperative market architecture, there exists a linear optimal
  stationary cooperative load scheduling strategy $u^c(x,d_2)$ given by:
  \begin{align}
    u^c(x,d_2)  =&-\underbrace{\frac{1}{1+\sqrt{1-q_2}}}_{a^c}x +
    \underbrace{\frac{1}{1+\frac{1}{\sqrt{1-q_2}}}}_{b^c}d_2
    %\nonumber\\ &
    + \underbrace{ \frac{q_1\mu_1+q_2\mu_2 \frac{1}{1+\sqrt{1-q_2}} }{1+\sqrt{1-q_2}}}_{g^c}
    \label{eq:2-linear-policy}
  \end{align}
%  which solves the Bellman equation in (\ref{eq:bellman2}).
\end{Prop}
\begin{Proof}
  Please refer to Appendix~\ref{prf:exist-coop-opt-L2}.
\end{Proof}

%\subsection{ Welfare Impacts}
\section{Welfare Impacts}
% With the linear equilibrium strategy and the linear optimal cooperative strategy given
% in (\ref{eq:2-linear-equil}) and (\ref{eq:2-linear-policy}), We examine the welfare
% impacts of realtime pricing and agent behaviors under different market architectures.

Given a linear strategy $u(x,d_2) = -ax + bd_2 + g$, ($a\in(0,1)$), we have the state
evolution dynamics:
\begin{align*}
  % x(t) = o_{(1,1)}(t) d_{1}(t) + o_{(2,2)}(t-1)\big( ax(t) + (1-b)d_{2}(t-1) - g)\big),
  x(t+1) = o_{(1,1)}(t+1) d_{1}(t+1) + o_{(2,2)}(t)\big(
  d_{2}(t) - u(x(t),d_2(t))
  \big) 
\end{align*}
which pins down the stationary distribution $\mc X$ of the aggregate backlog state
$\{x(t):t\in\mbb Z\}$ and $\mc U$ of the aggregate demand process $\{U(t):t\in\mbb Z\}$,
and it also determines the efficiency and risk performance.

% We move now to quantify the risk $R$ defined in (\ref{eq:def-risk}) for the $L=2$
% case. Note that given a linear strategy $u(x,d_2) = -ax + bd_2 + g$, the instantaneous
% aggregate demand is given by:
% \begin{align*}
%   U(t)& = x(t) + u(t)h_2(t)
%   \\&
%   = (1-a h_2(t))x(t) + h_2(t)(bd_2(t) +g)  
% \end{align*}

%%%%%%%%%%%%%%%%%%%%%%%%%%%%%%% this goes to proof of the prop%%%%%%%%%%%%%%%%%%%%%
%%%%%%%%%%%%%%%%%%%%%%%%%%%%%%%%%%%%%%%%%%%%%%%%%%%%%%%%%%%%%%%%%%%%%%%%%%%%%%%%%%%

Take expectation on both side of the aggregate backlog state dynamics, and we
obtain the first and second moment of $\mc X$ as follows:
\begin{align*}
  \mbb E[x(t)] = &\frac{q_1\mu_1 + q_2((1-b)\mu_2-g)}{1- q_2 a}
  \\
  \mbb E[x(t)^2] = &\frac{1}{1-q_2 a^2}\Big[
  q_1(\mu_1^2 + \sigma_1^2 ) + q_2\Big(((1-b)\mu_2-g)^2 + (1-b)^2\sigma_2^2\Big) + 2 q_1 q_2 \mu_1((1-b)\mu_2-g)
  \\
  &+ 2 \frac{a}{1-q_2 a}
  \Big(  q_2 ((1-b)\mu_2-g) + q_1 q_2 \mu_1 \Big) \Big( q_2 ((1-b)\mu_2-g) + q_1 \mu_1\Big)
  \Big]
\end{align*}
%\subsubsection{Efficiency}
Assuming that all type 2 agents adopt the same linear strategy $u(x,d_2) = -ax +
bd_2 + g$, market efficiency,  as defined in (\ref{eq:def-efficiency}), is given by:
\begin{align}
  W &=-\mbb E[U(t)^2]/2
  \nonumber\\ &
  = -\frac{1}{2}\Big(
  (1-q_2 + q_2(1-a)^2 ) \mbb E[x(t)^2]
  + 2q_2 (1-a)(b \mu_2 + g) \mbb E[x(t)]
  + q_2((b \mu_2 + g)^2 + b^2 \sigma_2^2)
  \Big)
  \nonumber
\end{align}
% where $\lambda=\mbb E[U(t)^2]$ is obtained by solving the following fixed point equation in
% $(\lambda, V(x))$:
% \begin{align}
%   \lambda + V(x) &= (1-q)(x^2 + \mathbb E_{d_1}[V(d_1) ])
%   % \nonumber\\ &
%   +q\mathbb E_{d_1,d_2}\left[(x+ u(x,d_2))^2   
%   +V(d_2 - u(x,d_2) + d_1) \right] \nonumber
% \end{align}
% Again, we conjecture and verify that the value function $V(x)$ is of quadratic form as $V(x) =
% A x^2 + B x$, and  $\lambda$ is given by:
% \begin{align*}
%   \lambda = &q_1((\mu_1^2 + \sigma_1^2)A + \mu_1 B)   
%   + q_2 ((\mu_2^2 + \sigma_2^2)(b^2 + (1-b)^2 A))
%   \\
%   &+ q_2 (g^2 + 2\mu_2 b g + A(g^2 -2g(q_1 \mu_1 + (1-b)\mu-2)) + B((1-b)\mu_2 - g)
%   )
% \end{align*}
% where 
% \begin{align*}
%   A = \frac{1+q_2(a^2-2a)}{1-q_2a^2}, \quad  
%   % B &= \frac{(Aa(1-b)+b(1-a))\mu_2 + Aa\mu_1 + (1-a-Aa)g}{(1-aq)/(2q)}
%   B = \frac{2\left(
%       a(1+ q_2 a^2 -2 q_2 a)(q_1 \mu_1 + \mu_2) + (1+ q_2 a^2 -2 a)(b\mu_2 + g)
%     \right)}
%   {(1-q_2 a)(1-q_2 a^2)}
% \end{align*}
% The average per period cost $\lambda$ can also be obtained in closed form.
% is given by:
% \begin{align}
%   \lambda = A\Big[&(\sigma_1^2+\mu_1^2) + q\Big((1-b)^2(\sigma_2^2+\mu_2^2) + g^2
%   %\nonumber\\ &
%   +2(1-b)\mu_1\mu_2 -2g(\mu_1 + (1-b)\mu_2) \Big)\Big]
%   \nonumber\\
%  + B\Big[&\mu_1 + q((1-b)\mu_2-g)\Big] + q(b^2(\sigma_2^2+\mu_2^2) + g^2 + 2bg\mu_2)
%   \label{eq:lm-calc}
% \end{align}
In particular, with the specific linear strategies $u^{nc}(\cdot,\cdot)$ and
$u^c(\cdot,\cdot)$, we can calculate the efficiency $W^{nc}$ and $W^c$ under the
non-cooperative and the cooperative market architectures.
The difference $\Delta = W^c - W^{nc}$ is positive and increasing in $q_2$, as well as
increasing in $\sigma_1^2$ and $\sigma_2^2$, the variance of the workload distributions.
The higher $q_2$ is, the larger efficiency loss of non-cooperative scheme will be, which
suggests that the cooperative load scheduling scheme becomes increasingly efficient as
the arrival rate of flexible loads increases.

However, the stationary distributions of the aggregate demand processes in
Figure~\ref{fig:pD} show that the cooperative scheme also thickens the right tail of the
outcome distribution, which extremely high aggregate demands are quantified as a higher
upper bound of risk in the following proposition.

\begin{Prop}[Upper bound on the risk $R$] 
  \label{prop:upperbound_R_L2}
  Suppose that the workload distribution $ D_i$ are Normal distributions $\mathcal
  N(\mu_i,\sigma_i^2)$ for $i =1,2$.  Given a linear strategy $u(x,d_2) = -ax + bd_2 +
  g$, ($a\in(0,1)$), which leads to a stationary aggregate backlog distribution $\mc X$,
  the probability of aggregate backlog exceeding $M$ is upper bounded by:
  \begin{align*}
    \Pr(x(t)>M)\le  \frac{1}{\sqrt{2\pi}m_1}e^{-\frac{m_1^2}{2}}
  \end{align*}
  where 
    \[
    m_1= \frac{M-\frac{\mu_1 + (1-b)\mu_2 - g}{1-a}}
    {\sqrt{\frac{\sigma_1^2 + (1-b)^2\sigma_2^2}{1-a^2}}}.
    \] 
  Moreover, if the following condition is satisfied:
  \begin{align}
    \frac{1-(1-a)^2}{1-a^2} > \frac{b^2}{\frac{\sigma_1^2}{\sigma_2^2} + (1-b)^2}
    \label{eq:ab_condition_bound}
  \end{align}
  the risk of aggregate backlog exceeding $M$ is upper bounded as follows:
  \begin{align}
    R = \Pr(U(t)>M) \le q\Pr(x(t)\ge M) + o(e^{-M}) \tx{ as } M\to\infty
    \label{eq:upperbound_X}
  \end{align}
\end{Prop}
\begin{Proof}
  Please refer to Appendix~\ref{prf:upperbound_R_L2}.
\end{Proof}
Note that both $\mbb E[x(t)]$ and $\mbb E[x(t)^2]$ are increasing in $a$, and decreasing
in $b$ and $g$. It is easy to verify that the stationary distribution of $x(t)$ induced
by the linear optimal cooperative strategy $u^c(\cdot, \cdot)$, has a larger mean and a
larger variance than that induced by the non-cooperative equilibrium strategy
$u^{nc}(\cdot, \cdot)$. In other words, the state of the aggregate backlog is more
volatile in the cooperative scheme.
%under the cooperative market architecture.
%
Also, when $\sigma_1=\sigma_2$, the cooperative market architecture leads to a higher
upper bound of risk than that under the non-cooperative market architecture. This is
consistent with the following simulation results where the cooperative scheme indeed
results in a higher risk than that in the non-cooperative scheme.
The interpretation of the condition in (\ref{eq:ab_condition_bound}) is that, when the
variance of flexible load realizations is sufficiently lower than that of the
uncontrollable load realizations, and when the coefficient $a$ is relatively large than
the coefficient $b$, the aggregate demand spikes are mostly contributed by the high
aggregate backlogs.
%}
\begin{Rem}[Interpretation of the coefficients]
  For a linear strategy $u(x,d_2) = -ax + bd_2 +g$ adopted by type 2 agents, the
  coefficient $a$ can be interpreted as the sensitivity to the aggregate backlog
  $x(t)$. A larger $a$ means that the strategy is more aggresive in absorbing the
  fluctuation of uncontrollable loads in the environment. Note that both $a^{nc}$ and
  $a^c$ are increasing in $q_2$.
  % = 1/\big(2(1+\sqrt{1-q/2})\big), = 1/\big(1+\sqrt{1-q}\big)
  Intuitively, with a higher type 2 arrival rate $q_2$, each type 2 agent is more
  aggresive in responding to $x(t)$ at their first period, anticipating that during the
  second period will arrive and respond to $x(t+1)$ in a similar aggresive way.
  Also note that for any arrival rate $q_2$, $a^{nc}< a^c$ always holds, and
  $a^c\in[0.5,1]$, $a^{nc}\in[0.25,0.2929]$, which means that type 2 agents alway respond
  less aggresively to the aggregate backlog $x(t)$ under the non-cooperative market
  architecuture. This can be understood as a result of their strategic behavior at
  equilibrium.
  Similarly, we can interpret the coefficient $b$ as the sensitivity to the realizations
  of $d_2(t)$. We also make the observations that $b^{nc} > b^c$, and both $b^{nc}$,
  $b^c$ are decreasing in $q_2$.
\end{Rem}
% \\
% Given the stationary distribution of $x(t)$, we have another way of calculating the second
% moment of aggregate demand $\lambda$:
% \begin{align*}
%   \lambda = (1-q)\sum_{k = 0}^{\infty}q^k\Big( q\mathbb E[\mathcal X_k^2] +
%   (1-q)\mathbb E[\big((1-a)\mathcal X_k + bd_2 + g\big)^2] \Big)
% \end{align*}
% where $\mathcal X_k$ and $d_2$ are independent.
% \textcolor{blue}{
% \begin{Rem}[Implication on Pricing]
%   So far, with given Markov pricing function and demand realizing based on load
%   shifting, the producer side dynamics is abstracted away from the system. However,
%   different mechanisms may give quite different demand evolutions and therefore ramping
%   cost. For example, the peak can induce an extremely high cost compared with small
%   fluctuations, then under the cooperative load shifting scheme producers welfare can
%   be largely harmed by the relatively high tail distribution. This motivates us to
%   consider more complicated pricing mechanism to explicitly incorporate the dynamics at
%   producer side.
% \end{Rem}
% }
% \begin{Rem}[Strategy design]
%   In order to meet a set of constraints on both variance and tail probability, we want
%   to design a mechanism to induce the linear stationary strategy $u(x,d_2) = -ax + bd_2
%   + e$ with the coefficients $a,b,e$ satisfying the following constraints:
%   \begin{align}
%     & \Pr(\mathcal X > M) \le z
%     \\
%     & W = \lambda - (\mu_1 + q\mu_2)^2 \le \overline{\lambda}, \quad\quad 0\le a\le 1
%   \end{align}
% \end{Rem}

% \subsection{Numerical results}
\section{Numerical results}
\label{sec:numerical-results}
In the following, we shall visualize the efficiency-risk tradeoffs. In particular, we
compare the stationary distribution of the aggregate demand process induced by four
different linear strategies. We have $u^c(\cdot, \cdot)$ from the cooperative scheme, and
$u^{nc}(\cdot,\cdot)$ from the non-cooperative scheme.  In addition, we define the
``naive load scheduling'' scheme to be $u^{naive}(x,d_2) = d_2/2$, in which case every
type 2 agent evenly splits his work load between his two periods, and define the ``no
load scheduling'' scheme to be $u^{no}(x,d_2) = d_2$, in which case every type 2 agent
completes his work load at his first period.
%\tc{blue}{
\begin{itemize}
  \item Figure~\ref{fig:pA_1} shows the efficiency performance, which is negatively
    proportional to the second order moment of the aggregate demand process, under the
    four strategies.
    We observe that as the arrival rate $q_2$ of type 2 agents increases, $\mbb
    E[U(t)^2]$
    % the second order moment
    increases for every strategy. This is mainly due to the increase in the workload. We
    also observe the efficiency loss of the non-cooperative load scheduling scheme when
    compared to the cooperative scheme for all arrival rates.
    % \item
    
    Figure~\ref{fig:pA_2} shows the variance of the aggregate demand process as the
    arrival rate $q_2$ increases from 0 to 1. The variance is contributed by the
    uncertainties from both the Bernoulli arrival process and the workload realizations,
    and effective load scheduling tends to attenuate the variance.  Since the uncertainty
    from the Bernoulli arrival process achieves its maximum at $q_2=1/2$, the variance
    versus the rate $q_2$ plots have the hump shape.  Also, we observe that the variance
    gap between the non-cooperative and the cooperative scheme increases as $q_2$
    increases. This indicates that the cooperative load scheduling becomes more powerful
    in terms of attenuating the aggregate demand variance when the arrival rate of
    flexible loads increases.
  \item Figure~\ref{fig:pB} compares the risk of spikes across the four strategies.
    % The stationary distribution of $U(t)$ is of mixed type due to the discrete Poisson
    % arrival and continuous distribution of load realizations.
    % \tc{red}{ For $M=???$, the risk as well as the upper bound of it are plotted for
    % the four strategies. We can observe a higher risk for the cooperative scheme.????
    % }    
    The 0.95-quantile of the stationary distribution of the aggregate demand process is
    plotted for each strategy. A higher 0.95-quantile is associated with a higher risk
    for some large constant $M$.
    The 0.95-quantile increases in $q_2$ mostly due to the heavier workload arrival.
    We also observe that as the arrival rate $q_2$ increases, risk increases most rapidly
    with the cooperative scheme, while the non-cooperative scheme gives the lowest risk
    for all $q_2$ and only slightly increases as the arrival rate increases.
    % \tc{red}{Add  the risk given by the bound, check whether the condition on $a, b$ is
    %   satisfied, whether the bound is tight.}
  \item Figure~\ref{fig:pC} shows the sample paths of the aggregate demand process under
    the non-cooperative and the cooperative market architecture.
    In Figure~\ref{fig:pC_1}, we observe that at a smaller time scale, the cooperative
    scheme can better smooth the aggregate demand process, which is consistent with the
    lower aggregate demand variance. However in Figure~\ref{fig:pC_2}, at a larger time
    scale, we can identify more demand spikes produced endogenously by the cooperative
    load scheduling scheme, corresponding to the higher risk of the cooperative scheme.
  \item Figure~\ref{fig:pD}
%    \tc{red}{actual distribution of the  Markov chain not available}
    plots the empirical distributions of the aggregate demand process in both linear
    scale in Figure~\ref{fig:pD_1} and in log scale in Figure~\ref{fig:pD_2}. We observe
    that under the cooperative market architecture, the distribution is more concentrated
    around the mean. However, associated with a higher risk, the distribution also has a
    heavier tail when compared to that in the non-cooperative scheme.

    Figure~\ref{fig:pE} shows the resulting aggregate demand stationary distribution of
    the cooperative and the non-cooperative load scheduling scheme under the non-negative
    demand constraint. We observes that it qualitatively resembles the corresponding
    distribution Figure~\ref{fig:pD} in most essential aspects.

  \end{itemize}

  \begin{Rem}[When do spikes occur]
    \label{rmk:where_spike}
    A better understanding the local interaction between agents with flexible loads also
    helps to discover the origin of endogenous risk, namely the triggers for demand
    spikes.
    On one hand, the instantaneous aggregate demand will be driven up when the workload
    realization $d_{l}(t)$ from either type of agent is extremely high, which corresponds
    to the rare events of the load arrival processes. We classify this type of spikes to
    be exogenous. Moreover, for bounded support of $\mb D_l$ and large enough constant
    $M$, the exogenous shocks do not directly contribute to the risk measure.
    On the other hand, an aggregate spike can also be produced endogenously
    % when a high backlog is built up and the agents with flexible loads, who can absorb
    % the endogenous shock, are absent in the market.    
    when there is a sudden absence of type 2 agent arrival after some consecutive periods
    during which type 2 agents continued to arrive, upon which event the accumulated high
    aggregate backlog at the deadline translates into a demand spike.

    When obtaining the risk upper bound in Proposition~\ref{prop:upperbound_R_L2}, we
    made use of the fact that
    % if the uncertainties of the uncontrollable loads are high enough,
    most of the spikes are produced endogenously.
    This observation is further confirmed by the conditional distributions of aggregate
    demand process in Figure~\ref{fig:whereSpikes1}.
   %\tc{red}{ calculate this instead of drawing the simulation!!! }
    We can see that the tail of the aggregate demand distribution is much larger
    conditional on that there is no type 2 agent arrival, and is much larger conditional
    on that the aggregate backlog is high.
    Intuitively, the more efficient a load scheduling strategy is, the more intense the
    backlog usage will be, and the resulting high backlog volatility leads to demand
    spikes.
\end{Rem}

We also point out that the tradeoffs we observed hold not only between the cooperative
and non-cooperative market architectures above, but also exist in a variety of
oligopolistic market architectures. Even when the agents can coordinate their actions and
are risk sensitive, so that large spikes are mitigated, the tradeoff still exists and is
shaped by different market achitectural properties.
In Appendix~\ref{sec:variationL2}, we provide two parameterized variations of the market
architectures, where the number of new arrival of each type can be great than 1, and
where the agents can be risk sensitive, seperately.
\chapter{General $L$ Analysis: Pricing}
\label{sec:pricing}

As illustrated in Figure~\ref{fig:sys_diag_1}, the agents who make their load scheduling
decisions can be viewed as a full state feedback controller, the control signal $\mb
u(t)$ is fed back to the plant and affects the system state evolution according to
(\ref{eq:state-evolve-gen1}) and (\ref{eq:state-evolve-gen2}), and the system output is
the aggregate demand process $\{U(t):t\in\mbb Z\}$.
In the case with $L=2$, even when the existing agents adopt a linear strategy, the system
dynamics is not linear since the type $2$ agents $(2,2)_t$ do not arrive at every period
$t$.
For a general $L$, the load scheduling strategy, which is determined under a specific
market architecture, does not form a linear time-invariant feedback controller.
% If a symmetric linear strategy is adopted by the agents that are present in the market,
% the system follows a jump linear dynamics and the mode switching events are determined by
% the Bernoulli arrival process.
% %
% As an example, the jump linear system diagram is shown in
% Figure~\ref{fig:jump_linear_combo}.
% More generally, for $L$ types of agents, the existence state is a $D_c$-dimensional
% vector, where $D_c = L(L+1)/2$.  There are in total $2^{D_c}$ possible existence
% states.  Assuming that linear strategies are adopted and the instantaneous demand from
% agent $(l,\tau)_t$ equals $u_{(l,\tau)}(t) = \mb F_{(l,\tau)}\mb x(t)$, each of the
% existence state corresponds to a linear mode. The Bernoulli arrival process together
% with the evolution of existence state lead to mode change.  \small even for $L=3$ is
% very messy Take $L=3$, and $q_l=q$ for all $l\in\mathcal L$ as an example, the
% transition among different mode is depicted in Figure~\ref{fig:jump_linear_combo} (b).
% It has $2^6 = 64$ states... very messy when the transition probabilities are specified.
%
%Moreover, for the general case,
The non-linearity as a result of the Bernoulli arrival processes complicates the
analysis, and there is no explicit solution to the equilibrium load scheduling strategy
under marginal cost pricing in both the cooperative and the non-cooperative schemes.
% \begin{figure}[h!]  \centering
%   \includegraphics[width = 0.6 \textwidth]{eps/jump_linear_combo.jpg}
%   \caption{Markov jumped linear system diagram, for $L=2$. There are two modes,
%     corresponding to the existence state of the new type 2 agent  $h_2 = 0$ and $h_2 = 1$.}
%   \label{fig:jump_linear_combo}
% \end{figure}

We realize that the main hurdle of analyzing the general $L$ case lies in the nonlinear
dynamics due to the intermittent agent arrivals.  To circumvent the problem we shall
introduce a modified system with surrogate performance measures, which resembles the
original system in the most essential ways and facilitates the analysis. The results
obtained in this LTI framework provide us some insights on the original non-linear system
dynamics and the efficiency-risk tradeoffs.
The two key modifications are listed and interpreted as follows:
\begin{Modif}
  The agent arrival rate $\mb q= \mb 1$, namely agents of all types arrive at every
  period, so that $\mb h(t) = \mb 1$ and $\mb o(t) = \mb 1$ for all $t$.
  \label{modif:mod1}
\end{Modif}
\begin{Modif}
  The second moment $\mbb E[z_2(t)^2]$ of the aggregate backlog process $z_2(t) = \mb e' \mb
  x(t)$ is used as a substitute measure for risk.
  \label{modif:mod2}
\end{Modif}
%\tc{red}{justification:}
Observations of the correlation between spikes and backlog, as well as the correlation
between spikes and the absence of flexible loads in Remark~\ref{rmk:where_spike} motivate
us to use the backlog volatility as a substitute measure for the risk.
% , which was previously defined as the tail probability of the aggregate demand process.
% In the following, we shall focus on the endogenously produced spikes, and the
% volatility of the aggregate backlog process, measured by its variance, will be used as
% a substitute risk measure.
%
Notice that there is no contradiction between the first two modifications.
We examine the case with $\mb q=\mb 1$, and the equilibrium strategy, as well as the
evolution of the backlog state in the regime of high arrival rate $\mb q$ will be similar
to the case of $\mb q=\mb 1$; however, absence of flexible loads still happens
exogenously with small probabilities, and upon which occurrences a high backlog is turned
into a demand spike. Therefore, the volatility of the aggregate backlog state is used as
a substitute measure for the risk of spikes.

We also normalize the load arrival process so that the average load realization
$\boldsymbol{\mu}$, the average backlog state $\mbb E[\mb x(t)]$, and the average demand
$\mbb E[\mb u(t)]$, are all zero vectors. We also assume the load arrival process $\{\mb
d(t):t\in\mbb Z\}$ is an i.i.d. process.
% since a constant shift of aggregate demand in every period amounts to a constant shift in
% aggregate backlog.
In summary, the system diagram of the modified system with linear dynamics is shown in
Figure~\ref{fig:sys_diag_2}. The performance measures are the variance of the two outputs:
\begin{align}
  %\mb z(t) =
  \left[
      \begin{array}{c}
        z_1(t)
        \\
        z_2(t)
      \end{array}
    \right] =\left[
      \begin{array}{c}
        \mb e' \mb u(t)
        \\
        \mb e' \mb x(t)
      \end{array}
    \right]. \nonumber
\end{align}

%------------------------------------------------------------------------------------

% In the previous section, we have modified our problem so that it fits into an LTI
% framework. Fixing the marginal cost pricing rule, we showed that any load scheduling
% scheme cannot achieve better outcomes than that specified by the Pareto front in
% Proposition \ref{prop:three-way}.
%
From the system operator's view, we are interested in how agent decision making is shaped
by the market architecture, and how the architecture should be designed so that the
desired agent behavior is induced.
Usually many of the market architectural properties, for example the degree of
cooperation and risk sensitivity of the agents, are given, and the system operator's only
freedom is to design the pricing rule.
In the following, we shall consider a non-cooperative setup within the LTI framework, and
examine the equilibrium load scheduling strategies under any linear pricing rule, which
will be decided by the system operator\footnote{ In Appendix~\ref{sec:conge-fee-degree},
  we study another example where the system operator regulates the architectural property
  of ``degree of cooperation'' by imposing a ``congestion fee'', which in effect works to
  internalize the payoff externalities.}.
In particular, we focus on the static linear pricing rules parameterized by coefficients
$\mb q_1$ and $\mb q_2$, in the form:
\begin{align}
  p(t) = \mb q_1'\mb x(t) + \mb q_2'\mb u(t).
  \label{eq:linear-pricing}
\end{align}
% \begin{enumerate}
% \item Static pricing, only depending on backlog $\mb x(t)$:
%   \[p(t) = \mb x(t)'\mb q\]  
% \item Static pricing, depending on $\mb x(t)$ and $\mb u(t)$:
%   \[p(t) = \mb x(t)'\mb q_1 + \mb u(t)' \mb q_2\]
% \item Dynamic pricing, with memory:
%   \[p(t) = \sum_{k=0}^{N-1} \mb x(t-k)'\mb q_{1, k} + \sum_{k=0}^{N-1} \mb u(t-k)'\mb
%   q_{2, k}\]
% \end{enumerate}
% where $\mb q$, $\mb q_1$, $\mb q_2$, and $\mb q_{i,k}$ are $D_c$ dimensional constant
% vectors.  is adaptive to the backlog state $\mb x(t)$ and instantaneous demand $\mb u(t)$
% We consider three cases as follows: In all cases, there exists a linear symmetric
% equilibrium. For the static cases , the linear load scheduling strategy is in the form of:% \[\mb u(t) = \mb F^* \mb x(t)\]
% and for the dynamic case, it is in the form of:
% \[\mb u(t) = \sum_{k = 0}^{N-1} \mb F^*_k \mb x(t-k)\]
The instantaneous demand decisions are made by individual agents under the deadline
constraints in a non-cooperative way.  We restrict ourselves to the linear symmetric
MPE, assuming that the load scheduling strategy, if exists, is in the following
form:
\begin{align}
  \mb u^*(t) = \mb F^* \mb x(t),
  \label{eq:consistency}
\end{align}
and we denote the $(l,\tau)$-th row of $\mb F^*$ by $\mb F_{(l,\tau)}^*\in\mbb
R^{D_c}$. By individual rationality, $u_{(l,\tau)}(t)$ is optimized by agent $(l,\tau)_t$
when he dynamically updates his load scheduling decision, forming the rational
expectation that all other agents are adopting the equilibrium linear strategy as in
(\ref{eq:consistency}). More specifically, apply the one-shot deviation principle at the
equilibrium and we have the optimal load scheduling decision $u_{(l,\tau)}^*(t)$ given
by:
\begin{align}
  \fa (l,\tau)\in\mc C, \tx{if } \tau >1,
  \quad &u_{(l,\tau)}^*(t) %= \mb F_{(l,\tau)}^* \mb x(t)
  = \arg\min_{u\in\mbb R} \left\{ \mb p(t) u + \mbb E[ \sum_{k=1}^{\tau-1}\mb p(t+k) \mb
    F_{(l,\tau-k)}^* \mb x(t+k) ] \right\}
  \label{eq:IR-cons}
  \\
  \tx{subject to: }
  & \mb u(t) = \mb F^*\mb x(t) + \mb e_{(l,\tau)} (u - \mb F^*_{(l,\tau)}\mb x(t))
  \nonumber\\
  & \mb u(t+k) = \mb F^*\mb x(t+k), \ \ \fa k>0
  \nonumber\\
  & p(i) = \mb q_1'\mb x(i) + \mb q_2'\mb u(i), \ \ \fa i
  \nonumber\\
  & \mb x(i+1) = \mb R_1\mb x(i) + \mb R_2 \mb d(i) - \mb R_1\mb u(i), \ \ \fa i
  %\ \ \tx{for  } k = 2,\dots, \tau-1,  
  % &\qquad \mb x(t+1) = \mb R_1(\mb I - \mb F^*)\mb x(t) + \mb R_2 \mb d(t) + \mb R_1 \mb
  % e_{(l,\tau)}(\mb F^*_{(l,\tau)} \mb x(t) - u),
  % \\
  % &\qquad \mb x(t+k) = \mb R_1(\mb I - \mb F^*) \mb x(t+k-1) + \mb R_2 \mb d(t+k-1) \ \ \tx{for
  % } k = 2,\dots, \tau-1,   
  \nonumber\\
  \tx{if } \tau =1, \quad & u_{(l,\tau)}^*(t) = \mb e_{(l,\tau)}\mb x(t), \nonumber
\end{align}
where $\mb e_{(l,\tau)}$ is a $D_c$ dimensional vector with the only non-zero element
being $1$ at the $(l,\tau)$-th position. Moreover, at the symmetric equilibrium the
rational expectation should be consistent with the best response strategy, namely
(\ref{eq:consistency}) should be satisfied.

A direct application of the principle of optimality to  (\ref{eq:IR-cons}) leads to 
% Simplify the first order conditions of the rationality constraints in we have:
\begin{align}
\mb F^* = f_{(\mb q_1, \mb q_2)} (\mb F^*).
\label{eq:fixed-point-F-f}
\end{align}
For given coefficients $\mb q_1, \mb q_2$, the $(l,\tau)$-th row of the mapping $f_{(\mb
  q_1, \mb q_2)}:\mbb R^{D_c\times D_c}\to \mbb R^{D_c\times D_c}$ is specified as follows:
\begin{align}
    f_{(\mb q_1, \mb q_2)}(\mb F)_{(l,\tau)} = \left\{
    \begin{array}[]{ll}
     \mb e_{(l,\tau)}'
      & \tx{if } \tau=1,
      \\
    \frac{\mb e_{(l,\tau)}'\mb R_1' \mb A_{(l,\tau)}
        \Big(\mb R_1(\mb I - \mb F) + \mb R_1\mb e_{(l,\tau)}\mb F_{(l,\tau)} \Big)
        - \Big(\mb q_1' + \mb q_2'(\mb F-\mb e_{(l,\tau)}\mb F_{(l,\tau)}) \Big)}
      {\mb e_{(l,\tau)}'\mb R_1' \mb A_{(l,\tau)} \mb R_1\mb e_{(l,\tau)}  + 2 \mb e_{(l,\tau)}'\mb
        q_2}
      &\tx{if  } \tau >1,
    \end{array}
  \right.
  \label{eq:fixed-point-f-def}
\end{align}
where
\[ 
\mb A_{(l,\tau)} =\sum_{k=1}^{\tau-1}
\Big(\big (\mb R_1(\mb I - \mb F)\big)^{k-1}\Big)'
\Big(
(\mb q_1 + \mb F'\mb q_2) \mb F_{(l,\tau-k)}
+ \mb F_{(l,\tau-k)}'(\mb q_1' + \mb q_2 '\mb F)
\Big)
  \Big(\big (\mb R_1(\mb I - \mb F) \big)^{k-1}\Big).
\]
The highly nonlinear mapping $f_{(\mb q_1,\mb q_2)}$ is not a contraction, and obtaining
the conditions on the parameters which guarantee the existence of a fixed point solution
to (\ref{eq:fixed-point-f-def}) is a challenging task. However, the equation still
provides a set of necessary conditions for the equilibrium strategies to satisfy.
An iteration algorithm with a carefully chosen initial guess will converge to such a
fixed point, and we will use numerical examples to show how the pricing parameter $\mb
q_1$ and $\mb q_2$ shift the equilibrium.

% Instead, we shall examine some numerical results in the following and draw some
% guidelines that the system operator should pay attention to when designning the pricing
% rule.
% %

% \tc{red}{ First, we consider the case with $L=3$. For the parameterized linear pricing
%   rule given in (\ref{eq:linear-pricing}), we plot the Pareto front of the aggregate
%   demand variance (-efficiency) and the aggregate backlog variance (risk) in
%   Figure~\ref{fig:pricing-front-12}. We examine the pricing rules that lead to outcomes
%   on the Pareto front, and we find that compared to the case of marginal cost pricing,
%   the system operator can optimize the pricing rule to improve efficiency and reduce risk
%   at the same time. Moreover, it shows that by putting a higher relative weight on the
%   most urgent backlog state, i.e., higher $\mb q_1(2)/\mb q_1(1)$ and $\mb q_1(3)/\mb
%   q_1(1)$, we can trade market efficiency with a lower risk along the Pareto front.
%  \\
%   Second, we examine the impact of the pricing rule sets on the location of the Pareto
%   front. In Figure~\ref{fig:pricing-front-3}, we compare the Pareto front induced by the
%   pricing rules in the form $p(t) = \mb q_1'\mb x(t) + \mb q_2'\mb u(t)$ and induced by
%   the pricing rules in the form $p(t) = \sum_{i=0}^{1}\mb q_{1,i}'\mb x(t-i) +
%   \sum_{i=0}^{1}\mb q_{2,i}'\mb u(t-i)$. We observe that by adopting a class of more
%   general pricing rules, the system operator is able to push the Pareto front.
%   }

\begin{Prop}[System operator's problem]
  \label{prop:operator-problem}

  Assume the system operator's utility function is increasing in efficiency and
  decreasing in risk, and in particular is linearly decreasing in both the volatility of
  aggregate demand and aggregate backlog as follows:
  \[J(\mbb E[z_1(t)^2], \mbb E[z_2(t)^2]) = -(\alpha_1\mbb E[z_1(t)^2] + \alpha_2 \mbb
  E[z_2(t)^2]).\]
  The system operator optimizes the parameterized pricing rule as defined in
  (\ref{eq:linear-pricing}) to maximizes its utility, and the optimal solution $(\mb
  q_1^*, \mb q_2^*)$ is given by solving the following problem:
  \begin{align}
    &\min_{\mb q_1, \mb q_2 \in\mbb R^{D_c}, \ \mb Q, \mb F\in \mbb R^{D_c\times D_c}}
    \alpha_1\mb e'\mb F_{(\mb q_1, \mb q_2)} \mb Q \mb F_{(\mb q_1, \mb q_2)}\mb e +
    \alpha_2\mb e'\mb Q\mb e
   \label{eq:system-operator-opt}
   \\
   &\tx{subject to: } \mb R_1 (\mb I-\mb F)\mb Q (\mb I-\mb F')\mb R_1'
   - \mb Q + \mb R_2\mb R_2'= \mb 0
   \label{eq:system-operator-opt-c1}\\
   &\qquad\qquad\ \mb F = f_{(\mb q_1, \mb q_2)}(\mb F)
   \label{eq:system-operator-opt-c2}
 \end{align}
 where $f_{(\mb q_1, \mb q_2)}$ is the mapping defined in (\ref{eq:fixed-point-f-def}).
\end{Prop}
\begin{Proof}
  Please refer to Appendix~\ref{prf:operator-problem}.
\end{Proof}

\chapter{General $L$ Analysis: fundamental tradeoff}
\label{sec:generalL}

In
% Section \ref{sec:pricing},
Chapter \ref{sec:pricing}, we have introduced the modified system, as well as evaluated
the MPE strategy and system performance in a non-cooperative setup under linear pricing
rules.  The following interesting questions naturally arise: are the equilibrium load
scheduling strategies in the non-cooperative setup optimal? If not, given the system
dynamics what are the optimal strategies? Does there exist a market architecture that
induces such optimal strategies?
%
% This section
This chapter is devoted to an examination of these questions.

Ideally, the desirable load scheduling should simultaneously maximize efficiency and
minimize risk, or equivalently in the modified setup, simultaneously suppress the
volatility of the two measured processes: $z_1(t)$ and $z_2(t)$.
A load scheduling strategy is defined to be Pareto optimal if there does not exist any
other strategy that makes the volatility of $z_1(t)$ smaller without making the
volatility of $z_2(t)$ larger, and a pair $(\mbb E[z_1(t)^2], \mbb E[z_2(t)^2])$ locates on
the Pareto front if it is achieved by a Pareto optimal strategy.
Unless the Pareto front trivially includes the point $(0,0)$, it dictates the limit of
the system performances with a downward sloping tradeoff curve between efficiency and
risk.
Also note that the concept of Pareto optimal load scheduling strategy does not rely on
market architecture specifications, in the sense that the system performance achievable
under any specific market architecture will be bounded by the Pareto front.  The Pareto
front thus serves as a benchmark to measure how far away a load scheduling strategy
induced by a specific market architecture is from the optimal strategies.

In order to neatly characterize the set of Pareto optimal load scheduling strategies, we
hereby introduce the third modification to the LTI system:
\begin{Modif}
  The deadline constraints, which require that all agents empty their backlogged load
  when they exit the market, are relaxed. Instead, we track the total load mismatch upon
  their deadline:
  \[z_3(t)= \mb e_L'(\mb x(t) - \mb u(t)),\] where $\mb e_L$ is a $D_c$-dimensional
  column vector with the first $L$ elements being ones and all others zero.  We define
  the second moment $\mbb E[z_3(t)^2]$ as the third performance measure. Note that the
  smaller the variance is, on average the more closely that deadline constraints are met,
  and when $\mbb E[z_3(t)^2]=0$, the deadline constraints are enforced.
  \label{modif:mod3}
\end{Modif}
Finally, after the three modifications, the system diagram with the inputs of load
arrival processes and the outputs $\mb z(t) = [z_1(t), z_2(t), z_3(t)]$ is shown in
Figure~\ref{fig:sys_diag_3}.
We generalize the tradeoff between efficiency and risk to a three-way tradeoff among
efficiency, risk, and load mismatch upon deadline, with the three-way Pareto optimal
strategies and the three-way Pareto front (a surface in the 3-dimensional space)
similarly defined.
Now we are able to cast the problem of finding Pareto optimal load scheduling strategy
into a $\mc H_2$ optimization problem with an unconstrained feedback controller, which
admits a convex characterization.

In order to trace out the Pareto front, we follow the standard multi-objective optimization
technique to scalarize the objective. Consider the weighted output process:
\[\mb z_\alpha(t) = [{\alpha_1} z_1(t), {\alpha_2} z_2(t), {\alpha_3}z_3(t)], \]
where $\alpha_i>0$, for $i=1,2,3$, and $\alpha_1^2 + \alpha_2^2 + \alpha_3^2 = 1$.  A
Pareto optimal load scheduling strategies minimize the $\mc H_2$ system norm for a given
weight $\alpha = (\alpha_1, \alpha_2, \alpha_3)$:
\begin{align*}
  &\min_{\{\mb u(t):t\in\mbb Z\}} \|\mb z_\alpha(t)\|_2^2\\
  &\text{subject to: } \mb x(t+1) = \mb R_1( \mb x(t)-\mb u(t)) +\mb R_2 \mb d(t)
\end{align*}

\begin{Prop} [Three-way Pareto front]
  \label{prop:three-way}
  \begin{enumerate}
  \item For given non-negative weight $\alpha = (\alpha_1, \alpha_2, \alpha_3)$, the
    corresponding Pareto optimal load scheduling strategy is static and linear in the
    system state $\mb x(t)$ as follows:
    \[
    \mb u(t) = \mb F_\alpha^* \mb x(t).
    \]
    where $\mb F_\alpha^* = \mb Q^*\mb P^{*-1}$, and ($\mb Q^*, \mb P^*$) is the unique
    solution to the following convex optimization problem:
    \begin{align*}
      &\min_{\mb Q, \mb P\in\mbb R^{D_c\times D_c}, \mb M\in\mbb R^{3\times 3}} \quad
      \rho
      \\
      \tx{ subject to:\ \ \ }
      &\mb Q> \mb 0,
      \\
      & \text{Trace}(\mb M)\le \rho,
      \\
      & \left[
        \begin{array}[c]{cc}
          \mb Q &  \left( \mb R_1\mb Q - \mb R_1\mb P \right) '
          \\
          \left( \mb R_1\mb Q - \mb R_1\mb P  \right) 
          & \mb Q - \mb R_2\mb R_2' 
        \end{array}
      \right] > \mb 0,
      \\
      & \left[
        \begin{array}[c]{cc}
          \mb Q &  \left(
            \mb C_1 \mb Q + \mb D_{12} \mb P \right) '
          \\
          \left(
            \mb C_1 \mb Q +\mb D_{12} \mb P \right) 
          & \mb M 
        \end{array}
      \right] > \mb 0.
    \end{align*}
    where
    \[\mb C_1 = [ 0\ \ {\alpha_2}\mb e \ \ {\alpha_3}\mb e_L]',
    \ \ \mb D_{12}=[{\alpha_1}\mb e \ \ \mb 0 \ \  -{\alpha_3}\mb e_L]'.\]
  \item Given a matrix $\mb F$ such that the feedback rule $\mb u(t) = \mb F\mb x(t)$
    stabilizes the system, the $\mathcal H_2$ norm of the three performance measures is
    given by:
    \begin{align*}
      \|z_1(t)\|_2^2 = \mb e'\mb F\mb Q_F \mb F^{'}\mb e ;
      \quad \|z_2(t)\|_2^2 = \mb e'\mb Q_F \mb e ;
      \quad \|z_3(t)\|_2^2 = (\mb e' - \mb e_L'\mb F)\mb Q_F (\mb e' - \mb e_L'\mb F)'
    \end{align*}
    where $\mb Q_F$ is the controllability Gramian given by solving the following
    equation:
  \[\mb R_1 (\mb I-\mb F)\mb Q_F (\mb I-\mb F')\mb R_1' - \mb Q_F + \mb R_2\mb R_2'= \mb
  0\]
  
  \end{enumerate}
\end{Prop}
% \tc{red}{if exists, should show it, otherwise, state rigorously ``if exists''.  Divide
%   this proposition into two parts, one optimal $\mb F$, one from equilibrium.  }
%\vspace{-40pt}  
\begin{Proof}
  Please refer to Appendix~\ref{prf:three-way}.
\end{Proof}
%\vspace{+15pt}

With different parameters of $\alpha=(\alpha_1, \alpha_2, \alpha_3)$, different Pareto
optimal solutions are produced, and we can trace out the Pareto front.  In particular,
the curve when restricting the three-way Parato front to the plane of $\|z_3(t)\|_2^2 =
\epsilon$ for $\epsilon\ll 1$ approaches the efficiency-risk tradeoff curve when the
deadline constraints are enforced, and the corresponding weight $\alpha$ satisfies
$\alpha_3/\alpha_1 \gg 1$ and $\alpha_3/\alpha_2 \gg 1$.

As an example, in Figure~\ref{fig:tradeoff3d_h2_L5_surf}, we plot the Pareto front for
the case with $L = 5$ to visualize the three-way tradeoff among the three system
performance measures. In Figure~\ref{fig:L_3d_1}, we observe that as we tighten the
constraint on load mismatch upon deadline, namely with a smaller $\beta_3$ in the
constraint $\|z_3(t)\|_2^2\le \beta_3$, the two-way Pareto front of efficiency and risk
shifts outward, which means that volatility of both aggregate demand and aggregate
backlog will increase. Similarly, as the constraint on the second performance measure
becomes tighter, namely with a smaller $\beta_2$ in the constraint $\|z_2(t)\|_2^2\le
\beta_2$, the Pareto front of the other two measures shifts outward.

The second part of Proposition \ref{prop:three-way} provides a way to evaluate the system
performance for any linear load scheduling strategies. In
Appendix~\ref{sec:classF-tradeoff}, we introduce some parameterized classes of heuristic
load scheduling strategies, the parameters of which reflect the market architectural
properties. Numerical results reveal how the tradeoffs among the three goals are shaped,
as well as how far they are away from the benchmark of the Pareto front characterized
above.

 \chapter{Conclusion}
\label{sec:conclusion}
In this paper, we proposed a framework to examine the welfare impacts of load scheduling
under different market architectures. We took the approach of modeling agent behavior
with dynamic oligopolistic games, and pointed out that different market architectures
induce different agent behaviors, which lead to a tradeoff between efficiency and risk at
the aggregate level.
Moreover, we provided a characterization of the efficiency-risk Pareto front. This is the
fundamental tradeoff limit for the system with load scheduling dynamics, in the sense
that the system performance induced by any market architecture is bounded by the front.

There are two directions of our future research. First, we would like to relax the
complete information assumption, and examine the model with a large number of coexisting
agents. This is the case in many real life applications including future electricity
market, where small entities that own generation powers are able to participate, and
system state is not globally available. Mean field game theory is a promising tool in
analyzing agent behavior in this dynamic stochastic game with a large number of
players. The interesting questions we want to address are: how do agents react to local
and systemic dynamics, and how is agent behavior shaped by the information structure?
Moreover, when in the limit the market becomes competitive, does similar efficiency-risk
tradeoff exist?

Secondly, we would like to look into the system operator's problem of optimizing the
pricing rule. In our current work, the system performance is determined by the
aggregation of autonomous agents's behavior, which relies on the pricing mechanism in an
intricate way, and there is no tractable way for the system operator to design the
pricing rule to induce the desired agent behavior. We are still exploring different
formulations which can give us some insights on the problem of pricing mechanism design.
More generally, realtime prices can be viewed as an endogenously generated payoff
relevant signal sent by the system operator to the agents, aiming to induce the rational
agents to respond to the signal in a desirable way. Another interesting question to ask
is: what are the signaling schemes in general that can incentivize the agents to behave
in certain ways?

\appendix
\chapter{Tables}
%\begin{table}
% \caption{Armadillos}
% \label{arm:table}
% \begin{center}
% \begin{tabular}{||l|l||}\hline
% Armadillos & are \\\hline
% our	   & friends \\\hline
% \end{tabular}
% \end{center}
%\end{table}

\begin{table}[h]
%  \centering
    \begin{tabular}{|l|l|}
      \hline
      $l\in\mathcal L$ &agent type\\
      $(l,\tau)_t\in\mathcal C$ & at time $t$, the type $l$ agent who will continue 
      to stay in the market for $\tau$ periods\\
      \hline
      $\mb d(t)\in\mbb R^{L}$  & new agent load realization  at time $t$\\ 
      $\mb h(t)\in\{0,1\}^L$ & new agent arrival event at time $t$\\
      $\mb x(t)\in\mbb R^{D_c}$ & backlog state\\
      $\mb o(t)\in\{0,1\}^{D_c}$ & existence state\\
      $\mb s(t)\in\mathcal S$ &  system state, $\mb s(t) = (\mb x(t), \mb z(t))$\\
      $\mb u(t)\in\mbb R^{D_c}$ & instantaneous  demand \\
      \hline      
      $p(t)$ & realtime price per unit resource\\      
      $U(t)$ & instantaneous aggregate demand \\
      \hline
      $\mb u^{nc}$ & symmetric Markov Perfect  Equilibrium (MPE) load scheudling strategy\\
      $\mb u^c$ & optimal stationary cooperative load scheduling strategy\\
      $W$ & efficiency\\
      $R$ & risk \\
      $B$ & robustness\\
      \hline
      % $u^{nc}(\cdot,\cdot)$& MPE strategy, $L=2$\\
      % $u^c(\cdot,\cdot)$& optimal cooperative strategy, $L=2$\\
      % $u^{\gamma}(\cdot,\cdot)$& MPE strategy with modified payoff function, $L=2$\\
      % $u^{nc,K}(\cdot,\cdot)$& MPE strategy for $K$ type 2 agents per period, $L=2$\\
      % $u^{nc,\theta}(\cdot,\cdot)$& optimal risk sensitive cooperative strategy,
      % $L=2$\\
      % $u^{c,\theta}(\cdot,\cdot)$& MPE strategy with risk sensitive agents, $L=2$\\
      % \hline
      % $\mathcal F_s$ &  set of feedback controllers that stabilize the
      % system\\
      % $\mathcal F_{DL}$ &  set of feedback controllers that enforce the
      % deadline constraints \\
      % $\mathcal F_{\alpha}$ & set of feedback controllers that assign total
      % weight
      % $\alpha$ on others' backlog state \\
      % \hline
\end{tabular}
\caption{Notations}
\label{table:notations}
\end{table}

\clearpage
\newpage

\chapter{Figures}

% \begin{figure}
% \vspace{2.4in}
% \caption{Armadillo slaying lawyer.}
% \label{arm:fig1}
% \end{figure}
% \clearpage
% \newpage

% \begin{figure}
% \vspace{2.4in}
% \caption{Armadillo eradicating national debt.}
% \label{arm:fig2}
% \end{figure}
% \clearpage
% \newpage

%Figures starts from here, format as below

%------------system diagram----------------------------------------------------
\vspace{+30pt}
\begin{figure}[h!]
  \centering
  \includegraphics[width = 0.4\textwidth]{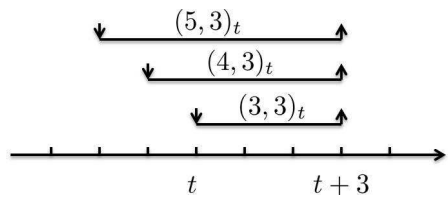}
  \caption{{Visualization of agent index $(l,\tau)_t$. For $L=5$, $\tau=3$, at time $t$
      there are at most 3 agents that will stay in the market for 3 periods. If they
      indeed arrive at the market, namely $h_3(t)=h_4(t-1) = h_5(t-2) = 1$, at time $t$
      they are indexed as $(5, 3)_t$, $(4,3)_t$ and $(3,3)_t$, seperately.}  }
  \label{fig:D1}
\end{figure}
\clearpage
 \newpage

\begin{figure}[h!]
  \centering
  \begin{subfigure}[b]{0.5\textwidth}
    \centering
    \includegraphics[width = 1 \textwidth]{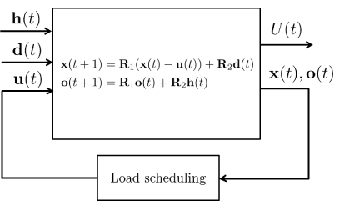}
    \caption{Orginal system dynamics with non-linear feedback
      controller.}
    \label{fig:sys_diag_1}
  \end{subfigure}

  \vspace{+20pt}
  \begin{subfigure}[b]{0.5\textwidth}
    \centering
    \includegraphics[width = 1 \textwidth]{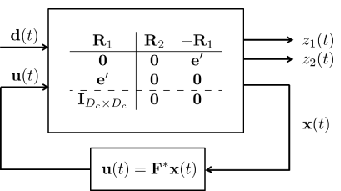}
    \caption{Linear time-invariant system formulation. There are two measurements:
      aggregate output process $z_1(t)$, and aggregate backlog process $z_2(t)$.  At
      equilibrium, load scheduling strategies of individual agents form a linear state
      feedback controller.}
    \label{fig:sys_diag_2}
  \end{subfigure}

  \vspace{+20pt}
  \begin{subfigure}[b]{0.5\textwidth}
    \centering
    \includegraphics[width = 1 \textwidth]{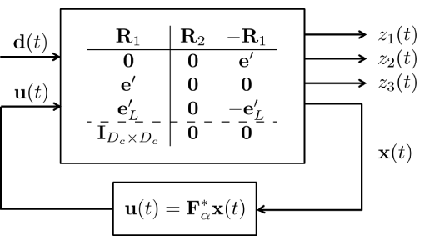}
    \caption{Linear time-invariant system formulation with relaxed deadline
      constraint. There are three measurements: aggregate output process $z_1(t)$,
      aggregate backlog process $z_2(t)$, and aggregate load mismatch upon deadline
      $z_3(t)$. $\mb u(t) = \mb F_\alpha^*\mb x(t)$ is a Pareto optimal load scheduling
      strategy profile.}
    \label{fig:sys_diag_3}
  \end{subfigure}

  \caption{System diagrams}
  \label{fig:sys_diag}
\end{figure}
\clearpage
\newpage

% -----------example of L=2 plots here-------------------------------------------
\begin{figure}[h!]
  \centering
  \begin{subfigure}[b]{0.5\textwidth}
    \centering
    \includegraphics[width = 1\textwidth]{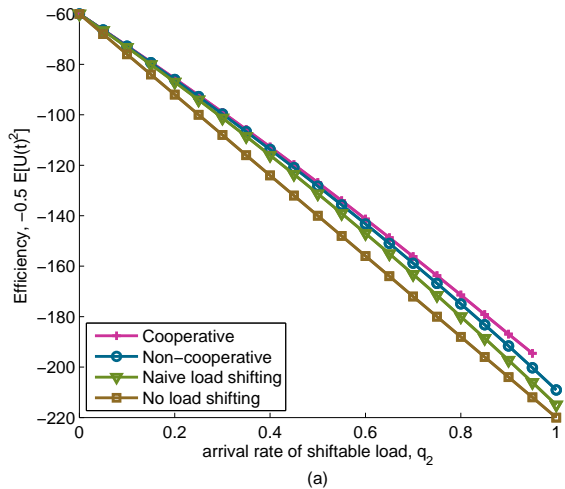}
    \caption{}
   \label{fig:pA_1}
  \end{subfigure}

  \vspace{+20pt}
  \begin{subfigure}[b]{0.5\textwidth}
    \centering
    \includegraphics[width = 1\textwidth]{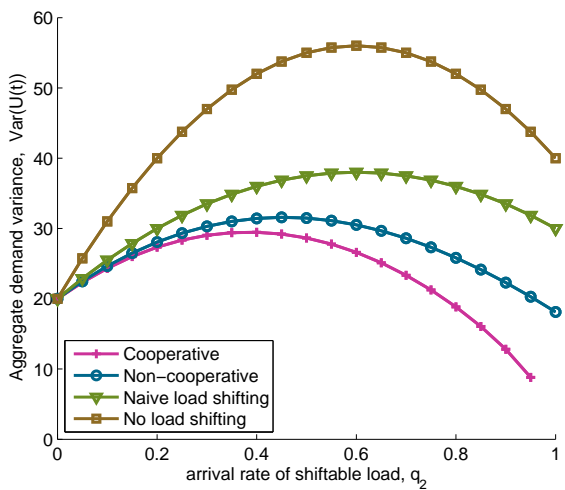}
    \caption{}
   \label{fig:pA_2}
  \end{subfigure}

  \caption{Market { efficiency} under different load scheduling schemes.  System
    parameters as follows: the number of agent types $L=2$; uncontrollable load Bernoulli
    arrival rate $q_1 = 1$; mean and variance of arrival load distribution $\mu_1 = \mu_2
    = 10$, $\sigma_1 = \sigma_2 = 11$. The cooperative load scheduling scheme leads to a
    lower aggregate consumption variance and thus a higher efficiency than that of the
    non-cooperative load scheduling scheme. This is known as the ``price of anarchy'' of
    strategic behavior in non-cooperative game.}
\end{figure}
\clearpage
\newpage

\begin{figure}[h!]
  \centering
  \includegraphics[width = 0.5\textwidth]{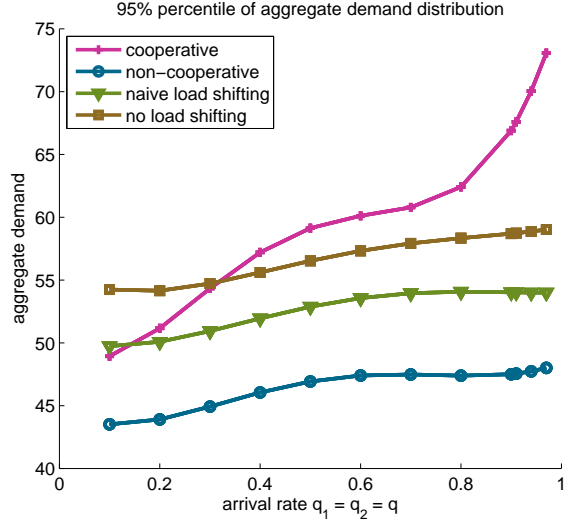}
  \caption{Risk under different load scheduling schemes.
    For any arrival rate $q$, the stationary distribution of the aggregate demand process
    has a larger tail under the cooperative market architecture than that under the
    non-cooperative market architecture.
    System parameters as follows: the number of agent types $L=2$; Bernoulli arrival rate
    $q_1 = q_2 = q$; mean and variance of arrival load distribution $\mu_1 = \mu_2 = 15$,
    $\sigma_1 = \sigma_2 = 4$.  }
  \label{fig:pB}
\end{figure}
\clearpage
\newpage

\begin{figure}[h!]

  \centering
  \begin{subfigure}[b]{0.6\textwidth}
    \centering
    \includegraphics[width = 1\textwidth]{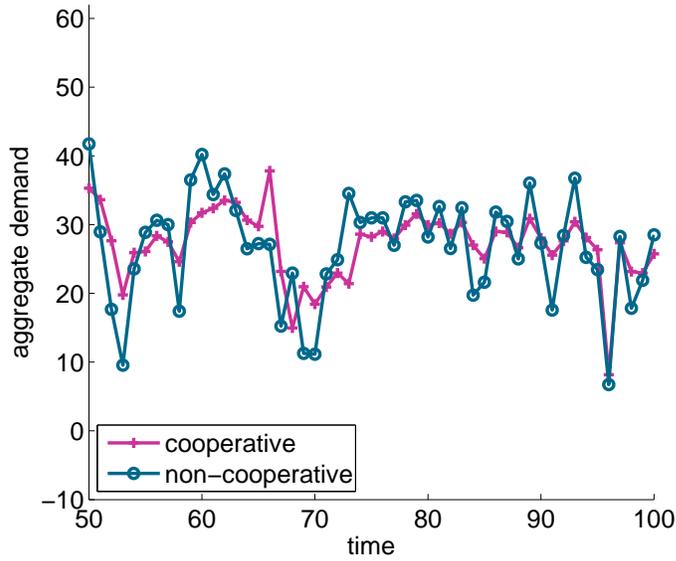}
    \caption{Short time scale }
    \label{fig:pC_1}
  \end{subfigure}

  \vspace{+20pt}
  \begin{subfigure}[b]{0.6\textwidth}
    \centering
    \includegraphics[width = 1\textwidth]{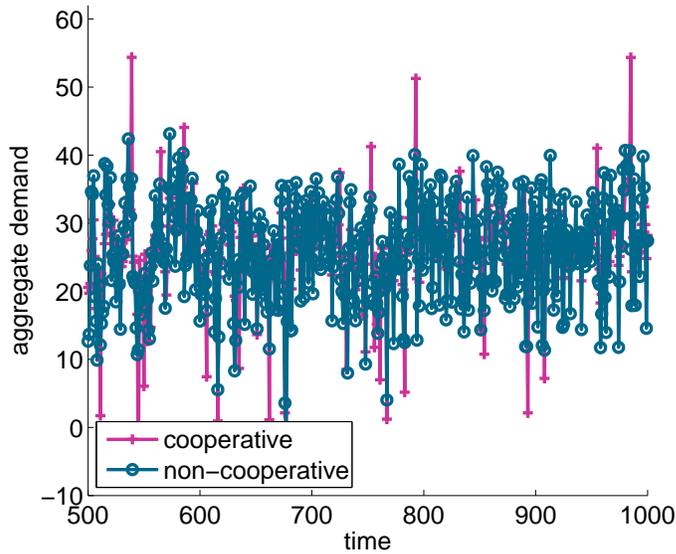}
    \caption{Large time scale }
    \label{fig:pC_2}
  \end{subfigure}

  \caption{Sample paths of the aggregate demand process under the cooperative and the
    noncooperative load scheduling schemes.
    At a smaller time scale, the cooperative load scheduling can better smooth out the
    aggregate demand process. However, at a larger time scale, there are more demand
    spikes produced endogenously by the cooperative load scheduling scheme. This is
    consistent with the observation of ``low variance, high tail probability'' of
    aggregate demand stationary distribution under the cooperative market architecture.
    System parameters as follows: the number of agent types $L=2$; Bernoulli arrival rate
    $q_1 = q_2 = 0.9$; mean and variance of arrival load distribution $\mu_1 = \mu_2 =
    15$, $\sigma_1 = \sigma_2 = 6$.}
    \label{fig:pC}
\end{figure}
\clearpage
\newpage

\begin{figure}[h!]
  \centering
  \begin{subfigure}[b]{0.5\textwidth}
    \centering
    \includegraphics[width = 1\textwidth]{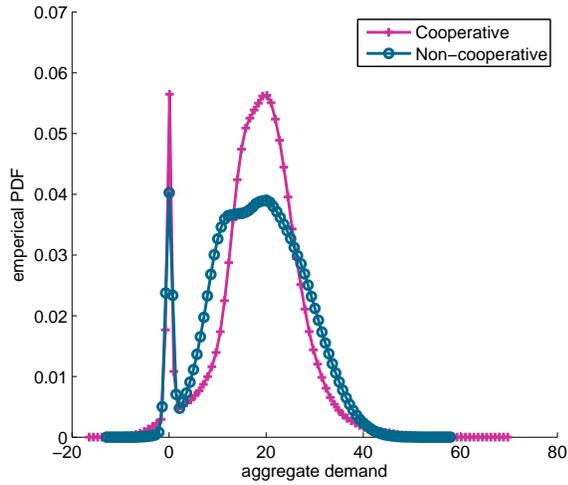}
    \caption{PDF }
    \label{fig:pD_1}
  \end{subfigure}

  \vspace{+20pt}
  \begin{subfigure}[b]{0.5\textwidth}
    \centering
    \includegraphics[width = 1\textwidth]{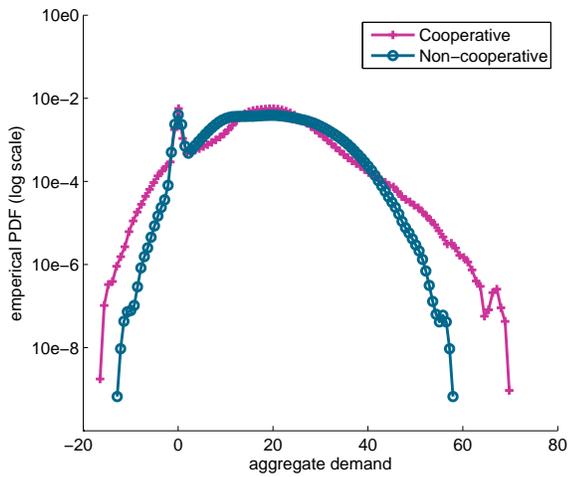}
    \caption{PDF (in log scale)}
    \label{fig:pD_2}
  \end{subfigure}

  \caption{Empirical distribution of the stationary aggregate demand process under the
    cooperative and the noncooperative load scheduling schemes.
    The stationary distribution under the non-cooperative market architecture is more
spread out but also has a smaller tail probability, while the distribution under the
cooperative market architecture is more concentrated around the mean but also has a
larger tail probability.
System parameters as follows: the number of agent types $L=2$; Bernoulli arrival rate
$q_1 = q_2 = 0.6$; mean and variance of arrival load distribution $\mu_1 = \mu_2 = 15$,
$\sigma_1 = \sigma_2 = 6$. }
  \label{fig:pD}

\end{figure}
\clearpage
\newpage

\begin{figure}[h!]
  \centering
  \begin{subfigure}[b]{0.5\textwidth}
    \centering
    \includegraphics[width = 1\textwidth]{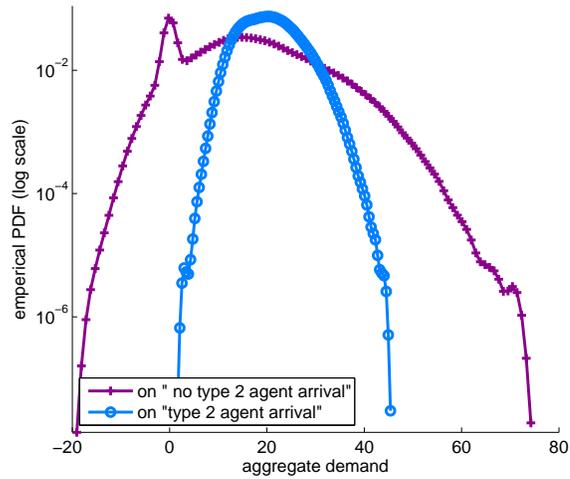}
    \caption{Aggregate demand distribution (log scale) conditional on whether flexible
      loads arrive or not}
  \end{subfigure}

  \vspace{+20pt}
  \begin{subfigure}[b]{0.5\textwidth}
    \centering
    \includegraphics[width = 1\textwidth]{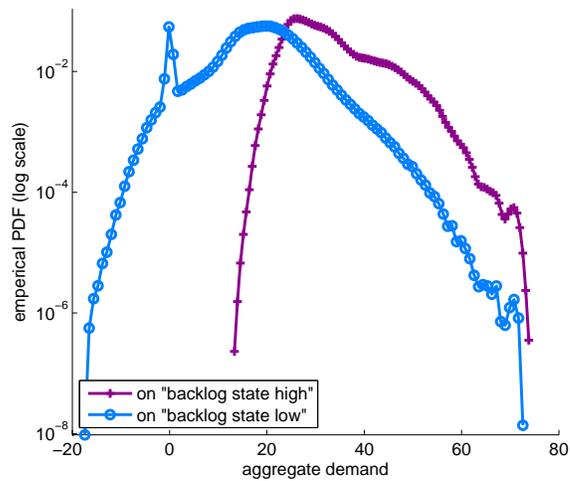}
    \caption{Aggregate demand distribution (log scale) conditional on whether aggregate
      backlog state is high or low}
  \end{subfigure}
  \caption{ Observations of when spikes happen. The extremely high aggregate demand
    (demand spikes) happen mostly when the flexible loads are absent and when the
    aggregate backlog state is high. Here we have $L=2$, $\mb D\sim \mc N(0,\mb I)$.}
  \label{fig:whereSpikes1}
\end{figure}
\clearpage
\newpage

% ------ 3-way tradeoff plots------------------------------------------
\begin{figure}[h!]
    \centering
    \includegraphics[width = 0.8 \textwidth]{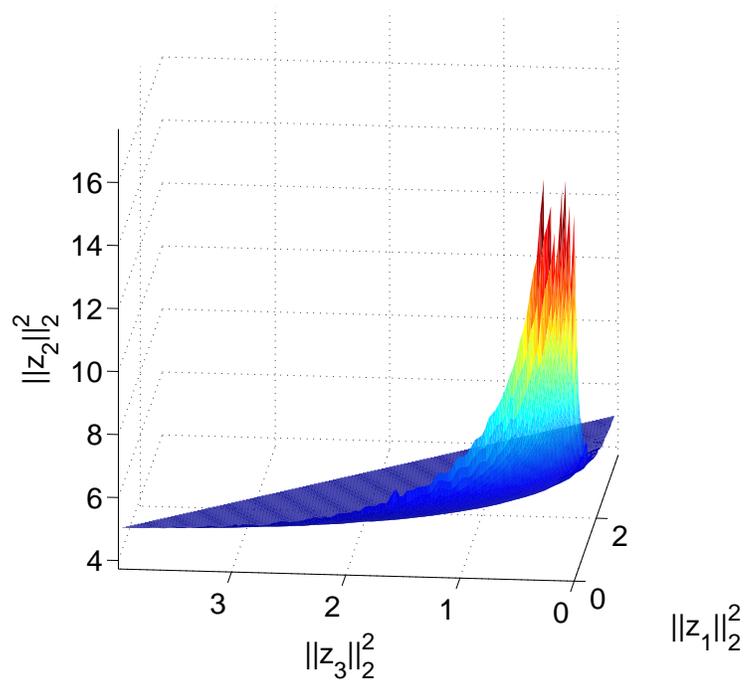}
    \caption{The three-way Pareto front. The three objectives are $\|z_1\|_2^2$,
      $\|z_2\|_2^2$, and $\|z_3\|_2^2$, which are the variance of the aggregate demand,
      the aggregate backlog, and the aggregate load mismatch upon deadline,
      correspondingly. Parameters: $L=5$.}
    \label{fig:tradeoff3d_h2_L5_surf}
  \end{figure}
\clearpage
\newpage

\begin{figure}[h!]
  \centering
  \begin{subfigure}[b]{0.7\textwidth}
    \centering
    \includegraphics[width = 1\textwidth]{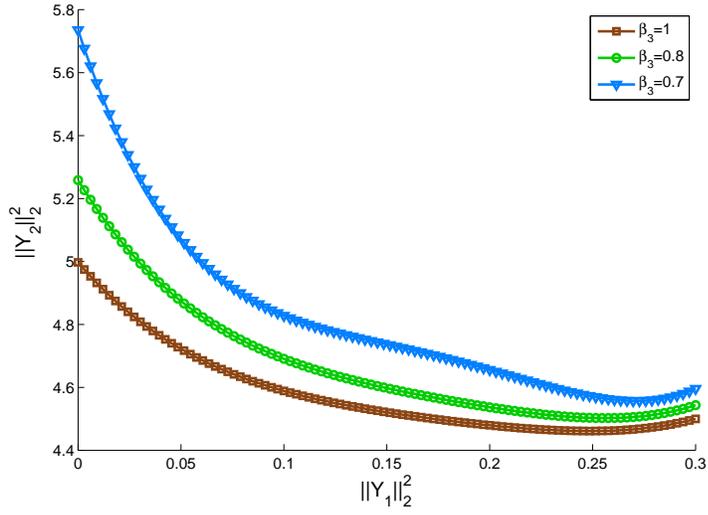}
    \caption{ When the constraint $\|z_3\|_2^2\le \beta_3$ is tightened, namely $\beta_3$
      decreases, the Pareto front of $\|z_1\|_2^2$ and $\|z_2\|_2^2$ shifts outward.  }
  \end{subfigure}

  \vspace{+20pt}
  \begin{subfigure}[b]{0.7\textwidth}
    \centering
    \includegraphics[width = 1\textwidth]{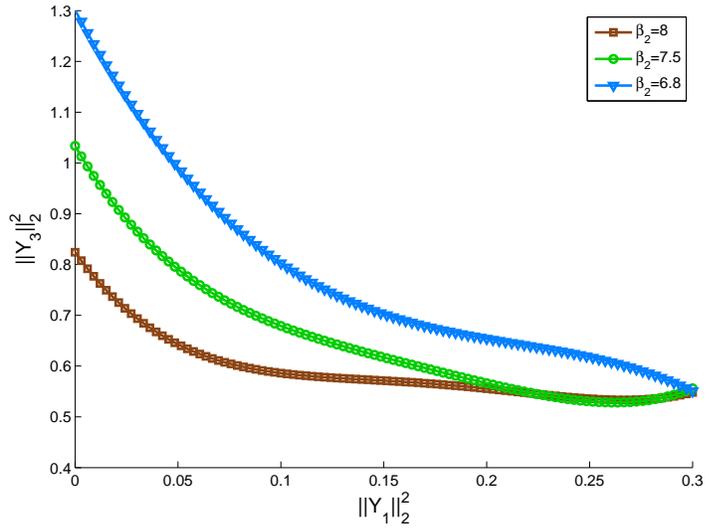}
    \caption{ When the constraint on $\|z_2\|_2^2\le \beta_2$ is tightened, namely
      $\beta_2$ decreases, the Pareto front of $\|z_1\|_2^2$ and $\|z_3\|_2^2$ shifts
      outward.}
  \end{subfigure}
  \caption{ Visualization of the three-way tradeoff Pareto front. The constraint on one
    performance measure affects the location of the tradeoff curve of the other two
    measures.  Parameters: $L=5$.}
    % , $\mb F\in\mathcal F_{\alpha=1}$}
  \label{fig:L_3d_1}
\end{figure}
\clearpage
\newpage

% \newpage

% % ===================================================================================
% % ==============================part B, not included in paper, but in thesiss========
% % ===================================================================================

\begin{figure}[h!]
  \centering

  \begin{subfigure}[b]{0.55\textwidth}
    \centering
    \includegraphics[width = 1\textwidth]{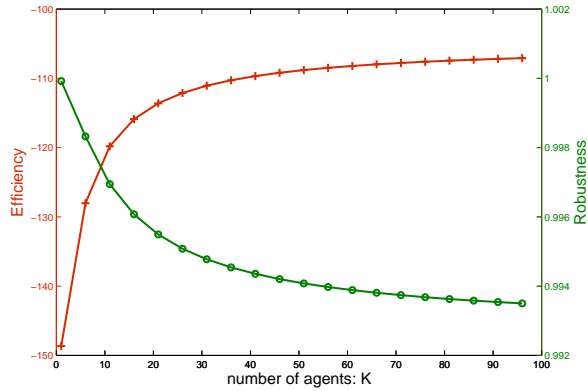}
    \caption{Market power leads to efficiency-risk tradeoff. As the number of agents $K$
      increases, individual's market power decreases, efficiency increases at the cost of
      robustness. }
    \label{fig:tradeoff3ways-1}
  \end{subfigure}

  \begin{subfigure}[b]{0.55\textwidth}
    \centering
    \includegraphics[width = 1\textwidth]{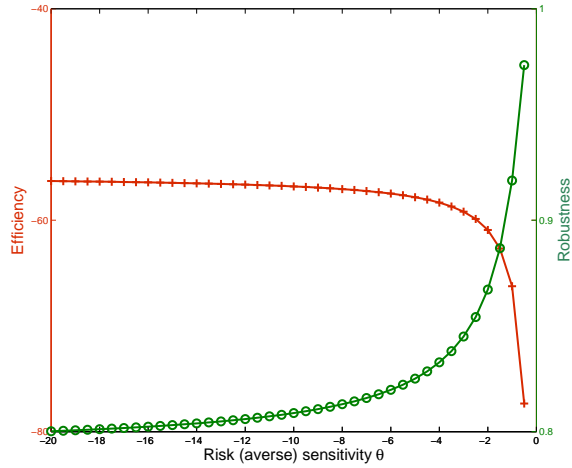}
    \caption{Risk sensitivity leads to efficiency-risk tradeoff. In the cooperative
      setup, as the maganitude of risk sensitivity $|\theta|$ increases, agent become
      more risk averse, and risk of spikes decreases. We observe a increase of robustness
      at the cost of a lower efficiency.}
    \label{fig:tradeoff3ways-2}
  \end{subfigure}

  \caption{Efficieny-risk tradeoffs as a result of different market architectural
    properties, in the case with $L=2$.}
\end{figure}
\clearpage
\newpage

\begin{figure}[h!]
  \centering
  % \vspace{+20pt}
    \includegraphics[width = 0.55\textwidth]{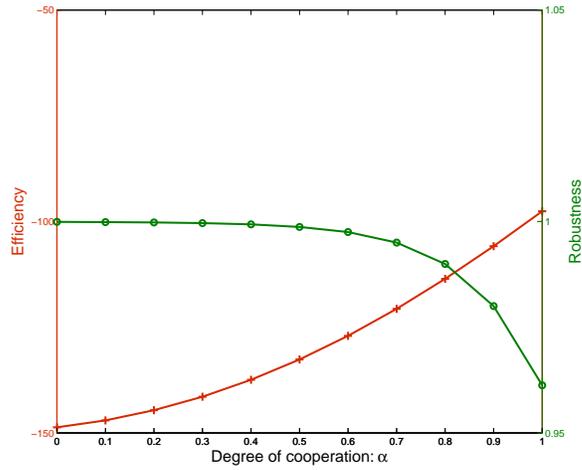}
    \caption{Degree of cooperation leads to efficiency-risk tradeoff. As the system
      operator increases the ``congestion fee'' by increasing $\alpha$, the payoff
      externality is reduced, and the degree of cooperation increases, leading to a
      higher efficiency and lower level of robustness.}
    \label{fig:tradeoff3ways-3}
\end{figure}

\begin{figure}[h!]
  \centering
  \begin{subfigure}[b]{0.5\textwidth}
    \centering
    \includegraphics[width = 1\textwidth]{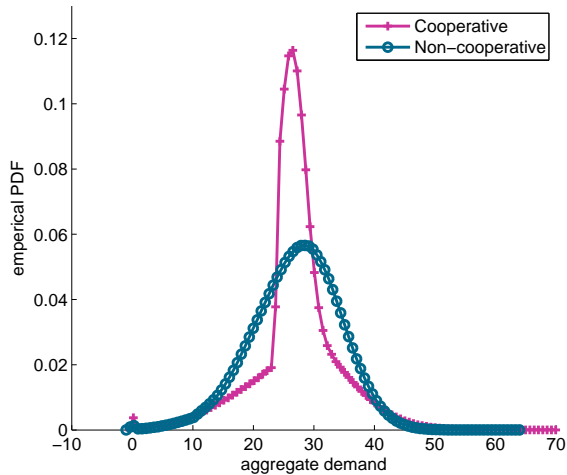}
    \caption{PDF }
    \label{fig:pE_1}
  \end{subfigure}

  \vspace{+20pt}
  \begin{subfigure}[b]{0.5\textwidth}
    \centering
    \includegraphics[width = 1\textwidth]{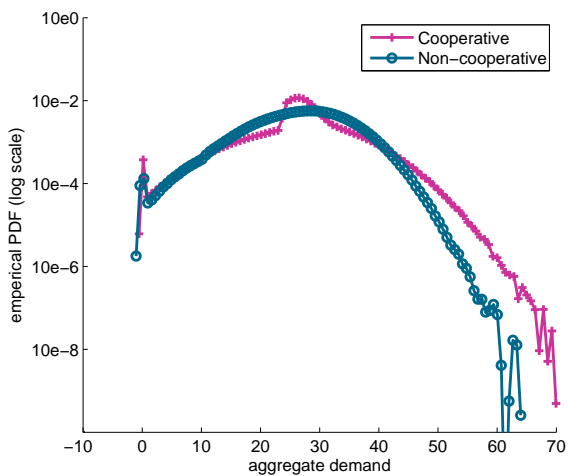}
    \caption{PDF (in log scale)}
    \label{fig:pE_2}
  \end{subfigure}

  \caption{ Empirical distribution of the stationary aggregate demand process under the
    constraint that the instantaneous demand from any agent is restricted to be
    non-negative. System parameters as follows: the number of agent types $L=2$;
    Bernoulli arrival rate $q_1 = q_2 = 0.6$; mean and variance of arrival load
    distribution $\mu_1 = \mu_2 = 15$, $\sigma_1 = \sigma_2 = 6$.  Under the bounded
    constraint. Observations similar to that for Figure~\ref{fig:pD} can be made.  }
    % Under the constraint that instantaneous demand from any agent cannot become
    % negative, we plot sample paths of the aggregate demand process.  System
    % parameters as follows:
    % the number of agent types $L=2$; Bernoulli arrival rate $q_1 = q_2 = 0.9$; mean
    % and variance % of arrival load distribution $\mu_1 = \mu_2 = 15$, $\sigma_1 =
    % \sigma_2 = 12$.  Observations
    % similar to those in Fig.~\ref{fig:pC} can be
    % made.  % Sample paths of the aggregate demand process, L=2 % Parameter setting as
    % below
    % $q_1 = q_2 = 0.9$, $\mu_1 = \mu_2 = 15$, $\sigma_1 = \sigma_2 = 12$ }
  \label{fig:pE}

\end{figure}
\clearpage
\newpage

\newpage

\chapter{Proofs}

% \section{Proof of Proposition~\ref{prop:exist-equil-general}}
% \label{prf:exist-equil-general}

\section{Proof of Proposition~\ref{prop:exist-linear-mpe}}
\label{prf:exist-linear-mpe} The result can be shown by first assuming that all other
type 2 agents adopt the conjectured linear strategy,
then verifying the first order conditions, and matching terms to obtain the coefficients
$a^{nc}$, $b^{nc}$, and $g^{nc}$. There is a unique root that leads to a dynamically
stable equilibrium.  % \tc{red}{ % specify the initial condition for the BR
sequence to converge.  % } % The uniqueness can be proven using the % iterative
expectation method and following the same argument as in % \cite{morris2002social}.
%\tc{red}{Follow the reference to show the steps of proof.}

\section{Proof of Proposition~\ref{prop:exist-coop-opt-L2}}
\label{prf:exist-coop-opt-L2}
We postulate the value function to be of quadratic form $ V^c(x) = A^cx^2 + B^cx$ , and
plug it in the R.H.S. of the Bellman equation.  Solve the minimization problem to get the
optimal strategy:
  \begin{align*}
    u^c(x,d_2) = &-\frac{1}{1+A^c}x + \frac{A^c}{1+A^c}d_2
    % \\&
    + \frac{A^c}{1+A^c}\mu_1 + \frac{B^c}{2(1+A^c)}
  \end{align*}
  Substituting back in the R.H.S., and matching terms on both sides yield the coefficients
  $A^c$, $B^c$, and the optimal per period cost $\lambda^c $:
  \begin{align*}
    \lambda^c =& A^c\sigma_1^2 + \frac{A^c}{1+A^c}q\sigma_2^2
    % \\&
    + \mu_1^2 + \frac{1+A^c-(A^c)^2}{1+A^c}q\mu_1^2 + 2q\mu_1\mu_2
  \end{align*}
  where
  \[ A^c = \sqrt{1-q}, \quad\quad B^c = 2(1-\sqrt{1-q})(\mu_2 + \mu_1).\]
  Therefore,
  $(\lambda^c, V^c(x) = A^cx^2 + B^c x)$ forms a solution to the Bellman equation, with the
  linear optimal stationary  strategy $u^c(x,d_2)$ in (\ref{eq:2-linear-policy}).

\section{Proof of Proposition~\ref{prop:upperbound_R_L2}}
\label{prf:upperbound_R_L2}
The stationary distribution of $U(t)$ is of mixed type due to the discrete Poisson arrival and
continuous distribution of load realizations.  Since the arrival process $\{h_2(t):t\in\mbb
Z\}$ of type 2 agents is exogenous, we first focus on the  distribution of
of the aggregate backlog process $\{x(t):t\in\mbb Z\}$.
When $|a|<1$, a  stationary distribution $\mathcal X$ exists and is characterized as follows:
\begin{align*}
  \mathcal X &= \mathcal X_k \ \ \text{with probability } q^k(1-q),\ \ (k=0,1,\cdots)
  \\
  \mathcal X_k& = \sum_{i=1}^{k} a^{i-1}\Big( D_{1,i} + (1-b) D_{2,i}\Big) + a^k D_{1,k} -
  \frac{1-a^k}{1-a} g
\end{align*}
where $\{ D_{1,i}:i\in\mbb Z^+\}$ and $\{ D_{2,i}:i\in\mbb Z^+\}$ are i.i.d. random
sequences respectively.  For every $k$, the mean and variance of the random variable
$\mathcal X_k$ are given by:
\begin{align*}
  \mathbb E[\mathcal X_k] = \frac{(1-a^{k+1})\mu_1 +
    (1-a^{k})((1-b)\mu_2 -g)}{1-a}
  % \\
  ,\qquad
  \text{Var}[\mathcal X_k] = \frac{(1-a^{2(k+1)})\sigma_1^2 +
    (1-a^{2k})(1-b)^2\sigma_2^2 }{1-a^2}.
  \end{align*}

% \[ R = \Pr(U(t)>M) \le \Pr(\mathcal X\ge M) \]
%   \begin{align}
%     \Pr&(U(t)> M)\le(1+\epsilon)\Pr(\mathcal X\ge M)\le\frac{ (1+\epsilon) }{\sqrt{2\pi}m}e^{-\frac{m^2}{2}}
%     \label{eq:upperbound_X}
%   \end{align}
%   where
%   {\small
%   \[
%   m= \left(M-\frac{\mu_1 + (1-b)\mu_2 - g}{1-a} \right)\Large/
%   \sqrt{\frac{\sigma_1^2 + (1-b)^2\sigma_2^2}{1-a^2}},
%   \]  }
%   \begin{align*}
%   m= \frac{M-\frac{\mu_1 + (1-b)\mu_2 - g}{1-a} }{\frac{\sigma_1^2 +
%       (1-b)^2\sigma_2^2}{1-a^2}},
%   \end{align*}
%   and $\Phi(\cdot)$ is the CDF of standard Normal distribution $\mathcal
%   N(0,1)$. Moreover, for a large enough $M$,

  Under the assumption that $ D_{1,i}, D_{2,i}\sim\mathcal N(\mu_i,\sigma_i^2), \
  i\in\{1,2\}$ the load distributions are normal, $\{\mathcal X_k:k\in\mbb Z^+\}$ are
  correlated normal random variables. Note that the mean and variance of $\mathcal X_k$
  are both increasing in $k$, we can upper bound the tail probability of $\mathcal X$ by
  the limiting distribution $\lim_{k\to\infty}\mathcal X_k$ as follows:
\begin{align*}
  \Pr(\mathcal X\ge M)
  % &= (1-q)\sum_{k=0}^{\infty}q^k\Pr(\mathcal X_k\ge M)
  % \nonumber\\
  & \le\Pr(\lim_{k\to\infty} \mathcal X_k\ge M)
\end{align*}
Since $\mbb E[\mbb E\lim_{k\to\infty}\mathcal X_k] = \lim_{k\to\infty}\mbb E[\mathcal
X_k]$ and $\lim_{k\to\infty}\text{Var}[\mathcal X_k]$, and it has normal distribution,
\begin{align*}
  \Pr(\mathcal X\ge M) = 1- \Phi\left(\frac{M-\frac{\mu_1 + (1-b)\mu_2 - e}{1-a}
    }{\frac{\sigma_1^2 + (1-b)^2\sigma_2^2}{1-a^2}}\right)
\end{align*}

\section{Proof of Proposition~\ref{prop:operator-problem}}
\label{prf:operator-problem}
The plant $G$ is given by
\begin{align*}
  G(s) = \left[
    \begin{array}[c]{c|cc}
      \mb A&\mb B_1 &\mb B_2
      \\
      \hline
      \mb C_1 & \mb 0& \mb D_{12}
      \\
      \mb I&\mb 0 &\mb 0
    \end{array}
  \right]
\end{align*}
where
\begin{align*}
  &\mb A = \mb R_1, \
  \mb B_1 = \mb R_2, \
  \mb B_2 =-\mb R_1 , \
  \\
  &\mb C_1 = [\mb 0\ \ \alpha_2\mb e\ \ \alpha_3\mb e_L]', \
  \mb D_{12} = [\alpha_1\mb e\ \ \mb 0\ \ -\alpha_3\mb e_L]'
\end{align*}
Consider the feedback gain $\mb F(s) = \mb D_K$ that stabilizes the system,
the closed loop system is given by
\begin{align*}
  \widetilde G(s) = \left[
    \begin{array}[c]{c|c}
      \mb A+\mb B_2\mb D_K &\mb B_1
      \\
      \hline
      \mb C_1 + \mb D_{12}\mb D_K &\mb 0
    \end{array}
  \right]
\end{align*}
$(\mb A+\mb B_2\mb D_K)$ is Hurwitz and $\|\widetilde G(s)\|< \rho$ if and only iff there
exists a symmetric matrix $\mb Q$ such that:
\begin{align}
  &(\mb A+\mb B_2\mb D_K)\mb Q (\mb A+\mb B_2\mb D_K)' -\mb Q + \mb B_1 \mb B_1^* < 0
  \label{eq:abd-1}
  \\
  &Trace(\mb C_1 + \mb D_{12}\mb D_K) \mb Q (\mb C_1 + \mb D_{12}\mb D_K)' < \rho
  \label{eq:abd-2}
\end{align}
Denote $\mb P = \mb D_K\mb Q$,  note that (\ref{eq:abd-1}), (\ref{eq:abd-2}) are equivalent to:
\begin{align*}
  &(\mb A\mb Q+\mb B_2\mb P)\mb Q^{-1} (\mb A\mb Q+\mb B_2\mb P)' -\mb Q + \mb B_1 \mb B_1^* < 0
  \\
  &Trace(\mb C_1\mb Q + \mb D_{12}\mb P) \mb Q^{-1} (\mb C_1\mb Q + \mb D_{12}\mb P)' < \rho
\end{align*}
Also, since trace is monotonic under matrix inequalities, we can finda matrix $\mb M$
such that $\mb M < \rho$ and
\begin{align*}
  (\mb C_1\mb Q + \mb D_{12}\mb P) \mb Q^{-1} (\mb C_1\mb Q + \mb D_{12}\mb P)' < \mb M
\end{align*}
Apply Schur's complement operation, we have that (\ref{eq:abd-1}), (\ref{eq:abd-2}) are
equivalent to the LMIs:
\begin{align*}
  &\left[
    \begin{array}[c]{cc}
      \mb Q & (\mb A\mb Q+\mb B_2\mb P)'
      \\
      \mb A\mb Q+\mb B_2\mb P & \mb Q -\mb B_1\mb B_1'
    \end{array}
  \right] >0,
  \ \
  \left[
    \begin{array}[c]{cc}
      \mb Q & (\mb C_1\mb Q+\mb D_{12}\mb P)'
      \\
      \mb C_1\mb Q+\mb D_{12}\mb P & \mb M
    \end{array}
  \right] >0.
\end{align*}
The Pareto optimal strategies can therefore be characterized by the convex optimization
problem to minimize $\rho$ with feedback gain $\mb F= \mb D_k = \mb Q\mb P^{-1}$.

\section{Proof of Proposition~\ref{prop:three-way}}
\label{prf:three-way}
In the non-cooperative setup, the system operator's optimization variables are the
pricing parameters $\mb q_1$ and $\mb q_2$. We have shown that for given $(\mb q_1, \mb
q_2)$ pair, at equilibrium agents' load scheduling strategy is the fixed point solution
to (\ref{eq:system-operator-opt-c2}). Under the assumption that load arrival process is a
i.i.d.  sequence, maximizing the system operator's utility is equivalent to minimizing
the $\mc H_2$ norm of the closed loop system, which is given by the objective in
(\ref{eq:system-operator-opt}), where $\mb Q$ is the controllability Gramian specified by
the Lyapunov equation in (\ref{eq:system-operator-opt-c1}).

\chapter{Supplementary Materials}
% (not to be included in the long version, but goes to the thesis.)

\section{Market Architecture Variations for $L=2$}
\label{sec:variationL2}
The tradeoffs we observed between coopeartive and non-cooperative schemes also exist in a
variety of oligopolistic market architectures. As an example, in this section, we provide
two parameterized variations of the market architectures, where the parameter $K$ allows
us to tune agents' market power; and parameter $\theta$ captures the risk sensitivity of
the agents.
In these two variations, strategies derived are still of linear forms, with the
coefficients as functions of $K$, and $\theta$, respectively.
In the following study of the case with $L=2$, our focus is the two period dynamics of
the representative type 2 agent. For notational convenience, we use $m$ and $m^+$ to
denote $m(t)$ and $m(t+1)$ for variables $m = x, u, d_1, d_2, p$.

\subsection{Number of Agents}
In the first variation, we adjust agents' market power by scaling the number of type 2
agents in the market.  We assume that when $h_2(t)=1$, $K$ homogeneous type 2 agents, all
denoted by $(2,2)_t$, simultaneously arrive at the market, each of them activates a job
with load requirement $d_2(t)/K$ and schedules his consumption over the two periods:
$(\nu^Kd_2(t)/K, \ (1-\nu^K)d_2(t)/K)$. Note that when $K =1$, it coincides with the case
of non-cooperative market architecture.
At equilibrium, each type 2 agent solves the problem:
\begin{align}
  &\nu^{K,*} = \arg\min_{\nu^K} \left\{ p\frac{d_2}{K}\nu^K + \mathbb
    E_{\{h_2^+,d_2^+,d_1^+\}}\left[ p^+\frac{d_2}{K}(1-\nu^K) \right] \right\}
\end{align}
where $x$ is the aggregate backlog state, and price is given by
\begin{align*}
  &p = x + \frac{d_2}{K} \big((K-1)\nu^{K,*} + \nu^K\big),
  \\
  &p^+ = x^+ + d_2^+\nu^{K,*}.
\end{align*}
Restricting to linear symmetric equilibra, we obtain an equilibrium strategy as follows:
\begin{align}
  u^K(x,d_2) =  \nu^{K,*}d_2& =
  - \underbrace{\frac{K}{K+1}\frac{1}{(1+\sqrt{1-\frac{K}{K+1}q_2})}}_{a^K} x
  + \underbrace{\frac{1}{1+\frac{1}{\sqrt{1-\frac{K}{K+1}q)2}}}}_{b^K} d_2
  \nonumber
  \\
  &+ \underbrace{\frac{\frac{K}{K+1}}{1+\sqrt{1-\frac{K}{K+1}q_2}}
    \left(
      q_1\mu_1 + q_2\mu_2\frac{1}{1+\sqrt{1-\frac{K}{K+1}q_2}}
      \right)
  }_{g^K}
  \label{eq:2-linear-equil-K}
\end{align}
\begin{Rem}[Limit when $K\to\infty$]
  Even though the agents with flexible loads are price anticipating and behave
  strategically, as the number of coexisting agents $K$ increases, their market power
  becomes diluted.  When $K$ increases to infinity, the equilibrium strategy converges to
  $u^c(x,d_2)$, the aggregate demand process converges to that of the cooperative
  scheme. The aggregate cost of all the type 2 agents is minimized, as well as the
  overall efficiency is maximized in the limit when $K\to\infty$.
  At a first glance, this convergence result contradicts to the Cournot limit theorem,
  which states that in a static partial equilibrium setting of quantity competition,
  profit maximizing firms become price-takers and the total profits decrease to zero when
  the number of firms increases to infinity \cite{green1978non}.
  However, our setup of the dynamic game is different from the Cournot competition in
  critical ways.  Under the marginal cost pricing and deadline constraints, the decisions
  $u(t)$ from groups of type 2 agents at consecutive periods are strategic complements,
  while within each group of $K$ identical type 2 agents, their decisions on first period
  consumption are strategic substitutes.
  Increasing $K$ leads to higher degree of within group competition which can potentially
  increase the group's cost in the sense of the Cournot limit theorem; however increasing
  $K$ also decreases each individual's market power and mitigates the cross group
  competition, which effect is dominant and overall results in a higher efficiency.
\end{Rem}
In Figure~\ref{fig:tradeoff3ways-1} we observe that as the market power decreases, market
efficiency increases while robustness decreases. In particular, when the agents become
price taking as $K\to\infty$, the first welfare theorem holds and market efficiency is
maximized, however the market is at the same time the least robust in terms of demand
spikes.

\subsection{Risk Sensitivity}
\label{sec:risk-sensitivity}
In the second variation, we consider the case where the agents are risk sensitive, and
examine the risk sensitive optimal load scheduling in a cooperative setup. In general,
risk averse agents tend to reduce the aggregate demand spikes, at the cost of a larger
variance of aggregate demand process.

We follow the Linear-Exponential-Quadratic-Gaussian (LEQG) framework in
\cite{whittle1990risk,hansen1995discounted} to study the risk sensitive optimal control.
Without loss of generality we assume $q_1=1$ and denote $q=q_2$.  Under the assumption
that the price is proportional to the instantaneous aggregate demand, the risk sensitive
objective function is defined recursively as follows:
\begin{align}
  c_t (x, d_2)=& q\mbb E_{d_2}\left[ (x+u)^2 - \frac{2\beta}{\theta}\log\mbb
    E_{d_1^+}[e^{-\frac{\theta}{2} c_{t+1}\left(d_2-u + d_1^+, d_2^+)\right) }] \right]
  \nonumber\\ & + (1-q)\left[ x^2 - \frac{2\beta}{\theta}\log\mbb
    E_{d_1^+}[e^{-\frac{\theta}{2} c_{t+1}\left(d_1^+, d_2^+)\right) }] \right]
\label{eq:risk-sensitive-objective}
\end{align}
We also assume the workload distributions Gaussian, namely $D_i\sim\mathcal
N(\mu_i,\sigma_i^2)$ for $i = 1,2$. The risk sensitivity is captured by the parameter
$\theta$. When $\theta<0$, the agents are risk averse, and when $\theta>0$, the agents
are risk loving.
Note that when $\theta<0$, the risk averse objective funciton in
(\ref{eq:risk-sensitive-objective}) imposes a larger disutility to large deviations from
the mean of $c_{t+1}(x^+, d_2^+)$, leading to higher penalties on the spikes than in the
risk neutral formulation. $\beta$ is the discount factor. As shown in
\cite{hansen1995discounted}, for $\theta<0$, there is a $\bar\beta(\theta)$
($0<\bar\beta<1$), such that for $\beta\le \bar\beta(\theta)$, a linear time invariant
optimal control policy exists.
% Moreover, $\bar\beta(\theta)$ is decreasing in the
% magnitude of $\theta$, namely $\bar\beta(\theta)$ is lower when the agents become more
% risk averse.
In our formulation, $\beta$ is chosen to be a small enough constant to ensure the
existence of a solution for the range of $\theta$ we consider.  Also note that when
$\theta\to 0$, $\bar\beta(\theta)\to 1$, the problem converges to the risk neutral case,
and the risk sensitive optimal cooperative strategy converges to that in
(\ref{eq:2-linear-policy}).
The risk sensitive optimal coopearative strategy minimizes the risk sensitive objective
function as follows:
\begin{align*}
  u^{c,\theta}(x(t), d_2(t)) = \arg\min_{u} c_t(x(t), d_2(t))
\end{align*}

\begin{Prop}
  For risk sensitivity $\theta\in\mbb R$, there exists a lower bound
  $\underline\beta(\theta)$ and an upper bound $\bar\beta(\theta)$, such that for
  $\underline\beta(\theta)\le\beta\le\bar\beta(\theta)$, there exists a risk sensitive
  optimal cooperative load scheduling strategy of linear form as follows:
  \begin{align}
    u^{c,\theta}(x,d_2) = -\underbrace{\frac{1}{1+r_3}}_{a^{c,\theta}} x
    + \underbrace{\frac{r_3}{1+r_3}}_{b^{c,\theta}} d_2
    + \underbrace{\frac{r_3(\mu_1 +\frac{r_1}{2r_2})}{1+r_3}}_{g^{c,\theta}}
    \label{eq:linear-policy-RS}
  \end{align}
  where the coefficients $r_i$ for $i=1,2,3$, are given by:
  \begin{align*}
    %\left\{
      %\begin{array}{ll}
        r_3 &= \frac{\beta r_2}{1+ \theta \sigma_1^2 r_2}
        \\
        r_2 &= \frac{(1-\beta-(1-q)\theta\sigma_1^2)
          \left(
            \sqrt{1+ \frac{4(1-q)(\beta+\theta\sigma_1^2)}{(1-\beta-(1-q)\theta\sigma_1^2)^2} } - 1
          \right)}
        {2(\beta + \theta\sigma_1^2)}
        \\
        r_1 &= \frac{2\beta r_2(1-r_2)(\mu_1+\mu_2)}{1+\theta\sigma_1^2 r_2 -\beta(1-r_2)}
      \end{align*}
\end{Prop}
% \begin{Proof}
% \end{Proof}
Note that under the cooperative market architecture, when the agents have a risk
sensitive objective function as above, the load scheduling strategy derived in
(\ref{eq:linear-policy-RS}) for $\theta\neq 0$ is different from the risk neutral optimal
strategy in (\ref{eq:2-linear-policy}). Nevertheless, system performance measures of
efficiency and robustness remain unchanged.  In Figure~\ref{fig:tradeoff3ways-2}, we
observe that when $\theta\le 0$ and as the magnitude of $\theta$ increases, the agents
become more risk averse, and the market efficiency decreases while the robustness
increases, and market efficiency achieves the maximum at $\theta = 0$.
% This is the case because maximizing the risk neurtral objective coincides with
% optimizing the market efficiency metric defined in (\ref{eq:def-eff-L2}).
%
Moreover, we notice that as the agents become risk loving for $\theta>0$, their objective
deviates from the market efficiency. Load scheduling produces more spikes at the
aggregate level, which have large negative impacts that bring down the overall efficiency
as well as increase endogenous risks.

\section{Numerical Study of Classes of Linear Load Scheduling Strategies}
\label{sec:classF-tradeoff}
% The following proposition formalizes the convexity
% as the tradeoff among the three goals.
% \tc{green}{
% \begin{Prop}[Three-way tradeoff with constrained strategies]
%   Let $(i,j,k)$ be a permutation of $\{1,2,3\}$, and define $z_k^*$ to be the solution to
%   the following optimization problem:
%   \begin{align*}
%     z_k^*(\beta_i,\beta_j) =\min_{\{\mb u(t):t\in\mbb Z\}} &\|z_k(t)\|_2^2\\
%     \text{subject to: }& \|z_{i}(t)\|_2\le \beta_i
%     \\
%     & \|z_{j}(t)\|_2\le \beta_j
%   \end{align*}
%   where $\mathcal F$ is a class of strategies that makes the optimization problem
%   feasible.  $z_k^*(\beta_i,\beta_j)$ is strictly decreasing in $\beta_i$ and $\beta_j$.
%   \label{prop:3way}
% \end{Prop}
% \begin{Proof}
%   Please refer to Appendix xxx.
% \end{Proof}
% }
%   \vspace{+15pt}

Through out this section, we restrict ourselves to linear load scheduling strategies:
\[ \mb u(\mb x(t)) = \mb F \mb x(t), \]
where $\mb F$ is a $D_c\times D_c$ dimensional matrix.

For general $L$, the Pareto front cannot be neatly characterized when there are
constraints on the feedback controller specified by $\mb u(t) = \mb F\mb x(t)$.
Next, we shall numerically examine how the market architectural properties, as reflected
by different constraints on $\mb F$, affect the location of the corresponding Pareto
front.

Intuitively, load scheduling should be operated according to the following principles:
firstly, with all other things being equal, an individual demands more resource when his
backlog is higher; secondly, when other agents' backlog states are high, he forms the
rational expectation that the instantaneous cost will be driven up, thus he consumes less
to avoid the high instantaneous price.
These are consistent with all the linear strategies we have examined for the case $L=2$,
which are of the form $u(x,d_2)= -a x + b d_2 + g$ where $a>0, b>0$.
Based on the above intuition, we consider the following constraint sets:
\begin{itemize}
% \item  $\mathcal F_s = \{\mb F\in \mbb R^{D_c\times D_c}:|\lambda_{max}(\mb
%   R_1(\mb I - \mb F)|<1\}$. This is the class of $\mb F$ that stabilize the system.
\item $\mathcal F_{DL} \triangleq \{\mb F\in \mbb R^{D_c\times D_c}: \mb F_{(l,1)}=\mb e_{(l,1)},
  \fa l\in\mathcal L\}$, where $\mb F_{(l,1)}$ is the row vector corresponding to the
  strategy of agent $(l,1)\in\mathcal C$, who meets his deadline, and $\mb e_{(l,1)}$ is
  a $D_c$ dimensional row vector with the $(l,1)$-th element being one and all others
  being zero.
  This is the constraint set in which deadline constraints  are enforced.
\item
  \begin{align*}
    \mathcal F_{\alpha} \triangleq \left\{ \mb F\in\mbb R^{D_c\times D_c} :
      \begin{array}{l}
        \mb F_{(l,\tau),(l,\tau)}=1,\ \ \fa (l,\tau)\in\mc C,
        \\
        \mb F_{(l,\tau),(l',\tau') } < 0, \ \ \fa (l',\tau')\neq (l,\tau) \in \mc C,
        \\
        \sum_{(l',\tau')\neq (l,\tau)}F_{(l,\tau),(l',\tau')} = \alpha.
      \end{array}
 \right\}
  \end{align*}
  for some $\alpha\le 1$. In this constraint set, an agent's instantaneous demand is
  negatively proportional to other agent's backlog state, with the sum being $\alpha$,
  and his demand is positively proportional to his own backlog with weight 1.  When
  $\alpha$ is small, the agent responds less aggresively to other agents, similar to the
  non-cooperative load scheduling that we observed in the case with $L=2$; when $\alpha$
  is high, the strategy is similar to the cooperative scheme.
\item
  \begin{align*}
\mathcal F_{BR,\delta}  = \left\{
\mb F\in\mbb R^{D_c\times D_c} :
\begin{array}{l}
  \mb F_{(l,1)} = \mb e_{(l,1)},\ \ \fa l\in\mathcal L
    \\
   \mb F_{(l,\tau),(l',\tau') } = \left\{
    \begin{array}{ll}
      1-\delta, & \text{ if } (l',\tau') = (l,\tau)
      \\
      -\frac{\delta}{D_c-1}, & \text{ if } (l',\tau')\neq (l,\tau)
    \end{array}
   \right.
\end{array}
\text{$\fa 1<\tau\le L$, $l\in\mc L$}
\right\}
\end{align*}
for some $\delta \in [0, 0.5]$\footnote{ The upperbound on $\delta$ is to ensure system
  stability for each $L\in\mathcal L$.}.  This is a parameterized class of {\sl boundedly
  rational load scheduling strategies} When the parameter $\delta$ is large, individual's
load scheduling decision is more sensitive to the other agents' backlog states and less
sensitive to his own backlog state.  This approximates the scenario when the market
architecture facillitates cooperation among agents.
\end{itemize}

The following corollary shows the impact of $\delta$ on aggregate demand volatility and
aggregate backlog volatility:
\begin{Prop}
  [Tradeoff of boundedly rational strategy] Assume that all agents adopt a boundedly
  rational load scheduling strategy $\mb u(t)=\mb F\mb x(t)$, where $\mb F\in\mc
  F_{BR,\delta}$. The aggregate demand volatility, measured by $\|z_1(t)\|_2^2$ is
  decreasing in $\delta$, and the backlog volatility, measured by $\|z_2(t)\|_2^2$ is
  increasing in $\delta$.
\end{Prop}
% \begin{proof}
% \end{proof}

Figure~\ref{fig:L_3d_1} shows how the total weight that an agent's linear strategy puts
on all other agents' backlog, i.e. $\alpha$, affects the Pareto front.  We observe that
as we decrease $\alpha$, the Pareto front shifts from the top left corner to the bottom
right corner, namely from high efficiency - high risk region to low efficiency - low risk
region.
This can be viewed as a generalization of our observation in the $L=2$ case.

% In the cooperative scheme, agents response more aggresively to other agents' backlog
% state, with a larger $b^{c}$ compared to $b^{nc}$ in the non-cooperative case;
% correspondingly, the cooperative scheme leads to higher efficiency and higher risk, while
% non-cooperative schemes leads to lower efficiency and lower risk.

\section{Congestion Fee and Degree of Cooperation}
\label{sec:conge-fee-degree}
% \tc{blue}{ Often the degree of cooperation is fixed in a given market setup, when both
% efficiency in terms of expected total payment and risk in terms of aggregate demand
% spikes are of the system operator's concern, the equilibrium solution may not be
% desirable. However, pricing mechanism provides another degree of freedom to locate the
% equilbrium solution on the efficiency-risk tradeoff curve.  }
In this example, the system operator can differentiate agents in the market. By imposing
a individual specific ``congestion fee'', the system operator is able to indirectly
adjust the level of cooperation of the market by changing agents' utility functions.

Recognizing that a key difference between the non-cooperative and the cooperative market
architecture is the payoff externality in the dynamic oligopolistic game, we introduce a
parameterized payoff function to attenuate the externality.
More specifically, for instantaneous price $p(t) = U(t)$, an agent pays for his own
demand at the price $p(t)$, and pays for a portion $\gamma (0\le \gamma\le 1)$ of the
instantaneous demand from all other agents at the same price $p(t)$.
For example, consider a type 2 agent with controllable load $d(t)$, on top of the total
cost $p(t)u(t) + p(t+1)(d(t)-u(t))$ for his consumption schedule, he also needs to pay
$\gamma p(t)x(t)$, and $\gamma p(t+1) (d_1(t+1) + h_2(t+1) u(t+1))$, during period $t$
and $(t+1)$\footnote{There should be an ex-ante money transfer from type 1 agents to
  type 2 agents in order to prevent type 2 agents from mimicing type 1 agents. However we
  do not explicitly calculate the amount of initial transfer for screening purpose, we
  shall instead focus on the equilibrium strategy of type 2 agents, and examine how the
  aggregate behavior affects the efficiency-risk tradeoffs at the macro level.}.  Note
that when $\gamma=0$, the induced strategy is the same as that under the original
non-cooperative market architecture; and when $\gamma=1$, the equilibrium strategy is
close to, though not equivalent to, the cooperative strategy where there is no payoff
externality among the agents.

With the level of payoff externality parameterized by $\gamma$, the
equilibrium load scheduling strategy is given by solving the following fixed point equation
\begin{align}
  u^{\gamma}(x,d_2)=\arg\min_u\Big\{ p(u+\gamma x)
  + \mbb E_{\{h_2^+, d_2^+, d_1^+\}}\left[ p^+(d_2-u
  + \gamma \left(d_1^+ + h_2^+u^{\gamma}(x^+, d_2^+))\right) \right] \Big\}
\end{align}
where $p(t) = U(t)$, and $x^+ = d_2-u + d_1^+$. The equilibrium strategy is given by:
\begin{align}
  u^{\gamma}(x,d_2) = -a^{\gamma}x + b^{\gamma}d_2 + g^{\gamma}
    \label{eq:2-linear-equil-gamma}
\end{align}
where the coefficients $a^{\gamma}$, $b^{\gamma}$, and $g^{\gamma}$ given by the following
system of equations:
\begin{align*}\left\{
    \begin{array}[]{l}
      \gamma q (a^{\gamma})^3 - (1+\gamma)q(a^{\gamma})^2 + 2a^{\gamma} - \frac{1+\gamma}{2} = 0
      \\
      b^{\gamma} = 1 - \frac{2a^{\gamma}}{1+\gamma}
      \\
      qg^{\gamma} =\frac{[(1-q)(1+\gamma) - q(2\gamma a^{\gamma} - 1-\gamma)(1-a^{\gamma}) ] \mu_1
        - q(2\gamma a^{\gamma} - 1-\gamma) b^{\gamma} \mu_2}
      {q(2\gamma a^{\gamma} - 1-\gamma) + \frac{1+\gamma}{a^{\gamma}}}
    \end{array}\right.
\end{align*}
We evaluate the market efficiency and the upper bound of risk for $\gamma\in[0,1]$. In
Figure~\ref{fig:tradeoff3ways-3}, we can observe the efficiency-risk tradeoff. As we
increase $\gamma$ from 0 to 1, the level of payoff externality decreases, and market
efficiency increases while robustness decreases, both monotonically.

\section{Example of state space model of LTI system for $L=3$}
\label{sec:example-state-space-L3}
As an example, for $L=3$
% the LTI system dynamics are given as follows:
% \begin{align*}
%   &\mb x(t+1) = \mb R_1(\mb x(t) - \mb u(t)) + \mb R_2 \mb d(t)
%   \\
%   & \mb u(t) = \mb F \mb x(t)
% \end{align*}
 and the two outputs that we measure are:
\begin{align*}
  & z_1(t) = [1\ 1\ 1\ 1\ 1\ 1\ ] \mb F\mb x(t)
  \\
  & z_2(t) = [1\ 1\ 1\ 1\ 1\ 1\ ] \mb x(t)
\end{align*}
The constant matrices $\mb R_1$, $\mb R_2$, and $\mb F\in\mc F_{BR,\delta}$ are given by:
  {\small
\begin{align*}
  &\mb R_1 = \left[
    \begin{array}{llllll}
      0    & 0    & 0    & 0    & 0    & 0\\
      0    & 0    & 0    & 1    & 0    & 0\\
      0    & 0    & 0    & 0    & 1    & 0\\
      0    & 0    & 0    & 0    & 0    & 0\\
      0    & 0    & 0    & 0    & 0    & 1\\
      0    & 0    & 0    & 0    & 0    & 0\\
    \end{array}
  \right],
  \ \ \ \
  \mb R_2 =  \left[
    \begin{array}{lll}
     1    & 0    & 0\\
     0    & 0    & 0\\
     0    & 0    & 0\\
     0    & 1    & 0\\
     0    & 0    & 0\\
     0    & 0    & 1\\
    \end{array}
  \right],
  \ \ \ \
  \\
  &\\
  &
   \mb F =  \left[
    \begin{array}{cccccc}
      1    & 0    & 0    & 0    & 0    & 0\\
      0    & 1    & 0    & 0    & 0    & 0\\
      0    & 0    & 1    & 0    & 0    & 0\\
      -\frac{\delta}{5}    & -\frac{\delta}{5}    & -\frac{\delta}{5}    & 1-\delta    & -\frac{\delta}{5}    & -\frac{\delta}{5}\\
      -\frac{\delta}{5}    & -\frac{\delta}{5}    & -\frac{\delta}{5}    & -\frac{\delta}{5}    & 1-\delta    & -\frac{\delta}{5}\\
      -\frac{\delta}{5}    & -\frac{\delta}{5}    & -\frac{\delta}{5}    & -\frac{\delta}{5}    & -\frac{\delta}{5}    & 1-\delta    \end{array}
  \right],
 \end{align*}
}
    
%% This defines the bibliography file (main.bib) and the bibliography style.
%% If you want to create a bibliography file by hand, change the contents of
%% this file to a `thebibliography' environment.  For more information 
%% see section 4.3 of the LaTeX manual.
\bibliography{smartGrid}
\bibliographystyle{plain}

\end{document}